\documentclass[a4paper,11pt]{article}
\pdfoutput=1 

\usepackage{jheppub} 

\usepackage[T1]{fontenc}
\usepackage{multirow,bbold,slashed}
 
\allowdisplaybreaks
 
\definecolor{nicered}{rgb}{0.7,0.1,0.1} 
\definecolor{nicegreen}{rgb}{0.1,0.5,0.1}
\definecolor{niceblue}{rgb}{0.0,0.1,0.7}
\hypersetup{colorlinks,citecolor=niceblue,linkcolor=niceblue,urlcolor=niceblue}

\usepackage[normalem]{ulem}

\def \bm#1{\mbox{\boldmath$#1$\unboldmath}}
\def \beq{\begin{equation}}
\def \eeq{\end{equation}}
\def \bea{\begin{eqnarray}}
\def \eea{\end{eqnarray}}

\title{A triple-threat challenge for $\bm{T}$ and $\bm{S}$ at two loops}

\author[a]{Ulrich Haisch,}
\author[a,b]{Sara Martellini}
\author[c]{and Ben~A.~Stefanek}

\affiliation[a]{Max Planck Institute for Physics, \\ Boltzmannstr.~8, 85748 Garching, Germany}
\affiliation[b]{Technische Universität M{\"u}nchen, Physik-Department, \\ James-Franck-Strasse 1, 85748 Garching, Germany}
\affiliation[c]{Instituto de F{\`i}sica Corpuscular (IFIC), Consejo Superior de Investigaciones Cient{\`i}ficas (CSIC) and Universitat de Val{\'e}ncia (UV), 46980 Valencia, Spain}

\emailAdd{haisch@mpp.mpg.de}
\emailAdd{saramart@mpp.mpg.de}
\emailAdd{bstefan@ific.uv.es}

\preprint{MPP-2026-211}

\abstract{In the~$Z_2$-symmetric real singlet-scalar extension of the Standard Model~(SM), the oblique parameters $T$ and $S$ first receive contributions at two loops, making their precise prediction a challenge. We take on this challenge with a triple threat. First, we perform an exact two-loop calculation in the full theory. Second, we derive the logarithmically enhanced terms from a general argument, according to which the singlet rescales the Higgs couplings to the electroweak gauge bosons and thereby spoils the cancellation of ultraviolet divergences between Higgs and Goldstone contributions. Third, we reproduce the results in the SM effective field theory, where the logarithms and mass-dependent terms stem from one-loop matrix elements of the dimension-six operator that modifies the Higgs kinetic term, and the remaining constants from two-loop matching. Our calculations provide a blueprint for new-physics scenarios in which the leading effects modify the Higgs couplings in a custodially symmetric way, so that custodial-symmetry violation, and hence $T$, arises only beyond one loop. Finally, we derive constraints from current and projected electroweak precision measurements and show that effects beyond dimension six matter for light singlets.}

\begin{document} 
\maketitle
\flushbottom

\section{Introduction}
\label{sec:introduction}

The Large Electron-Positron Collider~(LEP) and the SLAC Linear Collider~(SLC) transformed electroweak~(EW) physics into a precision science and established the Standard Model~(SM) as a quantitatively predictive theory. Operating at the $Z$ pole and above, these machines measured the properties of the $Z$ and $W$ bosons, the couplings of the fermions, and a variety of asymmetries with unprecedented accuracy. They thereby determined the EW parameters at the permille level, confirmed the gauge structure of the SM, and established the existence of three light neutrino species~\cite{ALEPH:2005ab}.

Decades later, EW precision observables~(EWPO) continue to define the precision frontier of particle physics and provide some of the most stringent indirect probes of physics beyond the SM~(BSM). Their precision is expected to improve substantially at future $e^+e^-$ colliders such as the Future Circular Collider~(FCC-ee)~\cite{FCC:2018evy} and the Circular Electron-Positron Collider~\cite{CEPCStudyGroup:2018ghi}. In many BSM scenarios, the dominant effects on EWPO are universal and can be captured by the oblique parameters $S$, $T$, and $U$ introduced by Peskin and Takeuchi~\cite{Peskin:1990zt,Peskin:1991sw}, which encode new-physics contributions to the self-energies of the EW gauge bosons. As~the experimental sensitivity improves, theoretical predictions for these parameters must keep pace, which in many models requires going beyond one loop.

A prime example is the Higgs portal, in which a real scalar singlet $\phi$ couples to the~SM exclusively through the Higgs doublet. If the singlet is odd under a $Z_2$ symmetry, it is stable and constitutes a dark matter~(DM) candidate~\cite{Silveira:1985rk,McDonald:1993ex,Burgess:2000yq,Patt:2006fw}. The $Z_2$ symmetry also forbids mixing between $\phi$ and the Higgs boson, so that the singlet affects EWPO neither at tree level nor at one loop. The leading contributions to $T$ and $S$ instead arise at two loops, and their computation is the central subject of this article. A first two-loop calculation of the oblique parameters in this model has been presented in~\cite{Englert:2020gcp}, with which our results do not agree, as discussed in~Section~\ref{sec:fulltwo}. Since collider searches lose much of their sensitivity once the singlet is too heavy to be produced in Higgs decays, EWPO provide a valuable complementary probe of this scenario.

We study the two-loop corrections to the oblique parameters in the $Z_2$-symmetric singlet-scalar model from three complementary angles. First, we perform an exact two-loop calculation in the full theory and derive compact analytic expressions for $T$ and $S$ in the limit of a heavy singlet. Second, we show that the logarithmically enhanced terms in these expressions follow from a general argument. The singlet modifies the Higgs couplings to the EW gauge bosons, which spoils the cancellation of ultraviolet~(UV) divergences between the Higgs and would-be Goldstone boson contributions and gives rise to infrared~(IR) logarithms~\cite{Barbieri:2007bh,Contino:2010rs,Panico:2015jxa}. Third, we reproduce these results within the SM effective field theory~(SMEFT)~\cite{Buchmuller:1985jz,Grzadkowski:2010es,Brivio:2017vri,Isidori:2023pyp}. There, the logarithms and all terms that depend on the EW masses arise from one-loop matrix elements of the dimension-six operator that modifies the normalisation of the Higgs kinetic term and thereby rescales the Higgs couplings to the~EW gauge bosons. The remaining constants correspond to two-loop matching corrections, which we compute explicitly. Together, these three approaches provide stringent cross-checks of our calculation and a transparent physical interpretation of the results. They also serve as a blueprint for other new-physics scenarios in which the leading effects modify the Higgs couplings in a custodially symmetric way, so that custodial-symmetry violation, and hence~$T$, arises only beyond one loop. Examples include extended Higgs-portal sectors, the custodial quadruplet model~\cite{Durieux:2022hbu,Maura:2024zxz,Maura:2025rcv}, and the two-Higgs-doublet model in the alignment limit with degenerate CP-odd and charged Higgs bosons~\cite{Bahl:2022jnx,Stylianou:2023xit}. Finally, we use our results to derive constraints on the singlet-scalar model from current and projected measurements of the $W$-boson mass, and we quantify the importance of effects beyond the dimension-six approximation for light singlets.

The remainder of this article is organised as follows. In~Section~\ref{sec:SSMandEFT}, we introduce the singlet-scalar model and its effective description. Section~\ref{sec:fulltwo} presents the full two-loop calculation of the oblique parameters, while Section~\ref{sec:IRlogs} explains the origin of the logarithmically enhanced terms. The SMEFT description of our results is developed in~Section~\ref{sec:SMEFT}, and the resulting constraints from the $W$-boson mass are discussed in~Section~\ref{sec:constraints}. We conclude in~Section~\ref{sec:conclusions}. The relevant loop functions are collected in~Appendix~\ref{app:loopfun}, the two-loop corrections to the $U$, $W$, and $Y$ parameters are presented in~Appendix~\ref{app:UWY}, the details of the two-loop SMEFT matching calculation are given in~Appendix~\ref{sec:SMEFTdetails}, and a dispersive representation of the gauge-boson self-energies is derived in~Appendix~\ref{app:dispersive}.

\section{The singlet-scalar model and its effective description}
\label{sec:SSMandEFT}

Throughout this article, we consider the SM extended by a real scalar singlet $\phi$ that interacts with the Higgs doublet $H$ via the Higgs portal. We impose a~$Z_2$ symmetry under which $\phi \to -\phi$, while all SM fields are even. This symmetry forbids all terms with odd powers of~$\phi$, such as a trilinear portal coupling $|H|^2 \hspace{0.5mm}\phi$ or a cubic self-interaction $\phi^3$, so that the Lagrangian takes the form
\beq 
{\cal L}_\phi = \frac{1}{2}\hspace{0.25mm}(\partial_\mu\phi)^2 - \frac{1}{2}\hspace{0.25mm}M^2\phi^2 - \frac{1}{2}\hspace{0.25mm}\kappa\hspace{0.25mm}|H|^2\phi^2\,,
\label{eq:Lphi}
\eeq
where the portal coupling $\kappa$ is real, as required by the hermiticity of the Lagrangian. We~have omitted the quartic self-coupling of $\phi$, since it does not contribute to the observables considered here at the order we are working. After EW symmetry breaking, the physical singlet mass is given by
\beq 
m_\phi^2 = M^2 + \frac{1}{2}\hspace{0.25mm}\kappa\hspace{0.125mm}v^2\,.
\label{eq:mphi}
\eeq

The~$Z_2$ symmetry renders the singlet stable, which makes it a DM candidate~\cite{Silveira:1985rk,McDonald:1993ex,Burgess:2000yq,Patt:2006fw}. If~$\phi$ is a thermal relic, direct-detection experiments severely constrain Higgs portals of the form~\eqref{eq:Lphi}. See, for instance,~\cite{Barger:2007im,Arcadi:2019lka,Ruhdorfer:2019utl,Lebedev:2021xey,Argyropoulos:2021sav}. For a non-thermal cosmological history, however, the singlet remains a viable DM candidate over a wide range of parameters~\cite{Hardy:2018bph}, which motivates probing the model at colliders. The sensitivity of collider searches depends primarily on~$m_\phi$. For $m_\phi < m_h/2$, with $m_h \simeq 125\,{\rm GeV}$ denoting the Higgs-boson mass, decays of the Higgs boson into invisible or undetected final states provide stringent constraints. See,~for example,~\cite{Djouadi:2011aa,Mambrini:2011ik,Djouadi:2012zc,Argyropoulos:2021sav}. Above this kinematic threshold, the singlet can be probed only through pair production in off-shell Higgs processes, such as vector-boson fusion~(VBF), $t \bar t h$ production, and gluon-gluon fusion~\cite{Matsumoto:2010bh,Kanemura:2011nm,Chacko:2013lna,Endo:2014cca,Curtin:2014jma,Craig:2014lda,Ko:2016xwd,Buttazzo:2018qqp,Ruhdorfer:2019utl,Heisig:2019vcj,Englert:2020gcp,Garcia-Abenza:2020xkk,Haisch:2021ugv,Detering:2026sgq,Chigusa:2026oek}, or through its virtual effects in processes such as $Zh$, double-Higgs, and $gg \to h^\ast \to ZZ$ production~\cite{Englert:2013tya,Craig:2013xia,He:2016sqr,Kanemura:2016lkz,Goncalves:2017iub,Goncalves:2018pkt,Englert:2019eyl,Haisch:2022rkm,Detering:2026sgq}. Although these channels have been studied for a broad range of hadron and lepton colliders, their sensitivity to singlets with $m_\phi > m_h/2$ turns out to be rather limited. This makes~EWPO an interesting complementary probe of the Higgs portal~\eqref{eq:Lphi}, and they are therefore the subject of this~work.

The~$Z_2$ symmetry also has important consequences for the effective field theory~(EFT) description of the model. Since $\phi$ couples to SM fields only in pairs, it cannot be exchanged at tree level, and integrating it out generates no tree-level contributions to the Wilson coefficients of the SMEFT. The leading effects arise at one loop, where the only dimension-six operators generated are~\cite{Ellis:2017jns,Jiang:2018pbd,Haisch:2020ahr,Cohen:2020fcu,Cohen:2020qvb,Carmona:2021xtq,Fuentes-Martin:2022jrf,Dittmaier:2026nnb}
\beq
Q_{H} = |H|^6\,, \qquad Q_{H\Box} = |H|^2 \Box |H|^2\,.
\label{eq:oneloopoperators}
\eeq
The operator~$Q_{H}$ only modifies the Higgs potential and plays no role in the following. The~operators that contribute to the Peskin-Takeuchi parameters $T$ and $S$ at tree level~\cite{Dawson:2019clf}, namely
\beq
Q_{HD} = \bigl(H^\dagger D^\mu H\bigr)^\ast \bigl(H^\dagger D_\mu H\bigr)\,, \qquad 
Q_{HW\!B} = \bigl(H^\dagger \sigma^a H\bigr) W_{\mu\nu}^a \hspace{0.125mm} B^{\mu\nu}\,,
\label{eq:operators}
\eeq
are instead first generated at two loops. The singlet therefore affects EWPO only at the two-loop level, which makes a precise theoretical treatment of these effects essential. Here $\Box = \partial^\mu\partial_\mu$ is the d'Alembertian, $D_\mu$ is the $SU(2)_L\times U(1)_Y$ covariant derivative, $B_{\mu\nu}$ and $W_{\mu\nu}^a$ are the $U(1)_Y$ and $SU(2)_L$ field-strength tensors, and $\sigma^a$ are the Pauli matrices. All~conventions follow the Warsaw basis~\cite{Grzadkowski:2010es}.

At the matching scale $m_\phi$, the one-loop Wilson coefficient of $Q_{H\Box}$ is given by~\cite{Ellis:2017jns,Jiang:2018pbd,Haisch:2020ahr,Cohen:2020fcu,Cohen:2020qvb,Carmona:2021xtq,Fuentes-Martin:2022jrf,Dittmaier:2026nnb}
\beq
C_{H \Box}^{(1)} (m_\phi) = -\frac{1}{16 \hspace{0.125mm} \pi^2} \hspace{0.125mm} \frac{\kappa^2}{24 \hspace{0.25mm} m_\phi^2}\,, 
\label{eq:high}
\eeq
where here and in the following the superscript indicates the loop order at which a Wilson coefficient is generated. The difference between $M$ and $m_\phi$ in~\eqref{eq:high} corresponds to terms of higher order in $v^2/m_\phi^2$, which lie beyond the dimension-six approximation.

\section{Two-loop oblique corrections in the full theory}
\label{sec:fulltwo}

Having set up the model and its EFT description, we now turn to the full two-loop calculation of the oblique corrections. They are determined by the transverse gauge-boson self-energies $\Pi_{VV^\prime}(q^2)$ with $VV^\prime = WW, ZZ, \gamma\gamma, \gamma Z$, defined through the decomposition
\beq
\Pi_{VV^\prime}^{\mu\nu}(q^2) = g^{\mu\nu}\hspace{0.25mm}\Pi_{VV^\prime}(q^2) + \ldots \,,
\label{eq:defPi}
\eeq
where $q^2$ denotes the invariant momentum transfer squared and the ellipsis represents terms proportional to $q^\mu q^\nu$, which vanish when contracted with conserved currents and thus do not contribute to the observables considered here. Expanding the transverse self-energies about $q^2=0$,
\beq
\Pi_{VV^\prime}(q^2) = \Pi_{VV^\prime}(0) + q^2\,\Pi^\prime_{VV^\prime}(0) + \mathcal{O}(q^4) \,,
\label{eq:Piexp}
\eeq
with the prime denoting differentiation with respect to $q^2$, the oblique parameters read~\cite{Peskin:1990zt,Peskin:1991sw}
\begin{gather}
T = \frac{1}{\alpha}\left[ \frac{\Pi_{WW}(0)}{m_W^2} - \frac{\Pi_{ZZ}(0)}{m_Z^2} \right] \,, \label{eq:defT} \\[2mm]
S = \frac{4 \hspace{0.125mm} s_w^2 \hspace{0.125mm} c_w^2}{\alpha}\left[ \Pi^\prime_{ZZ}(0) - \frac{c_w^2 - s_w^2}{s_w \hspace{0.125mm} c_w}\,\Pi^\prime_{\gamma Z}(0) - \Pi^\prime_{\gamma\gamma}(0) \right] \,, \label{eq:defS}
\end{gather}
where $\alpha$ is the electromagnetic coupling, $s_w$ and $c_w$ denote the sine and cosine of the weak mixing angle, and $m_W$ and $m_Z$ are the $W$- and $Z$-boson masses.

\begin{figure}[t!]
\begin{center}
\includegraphics[width=0.95\textwidth]{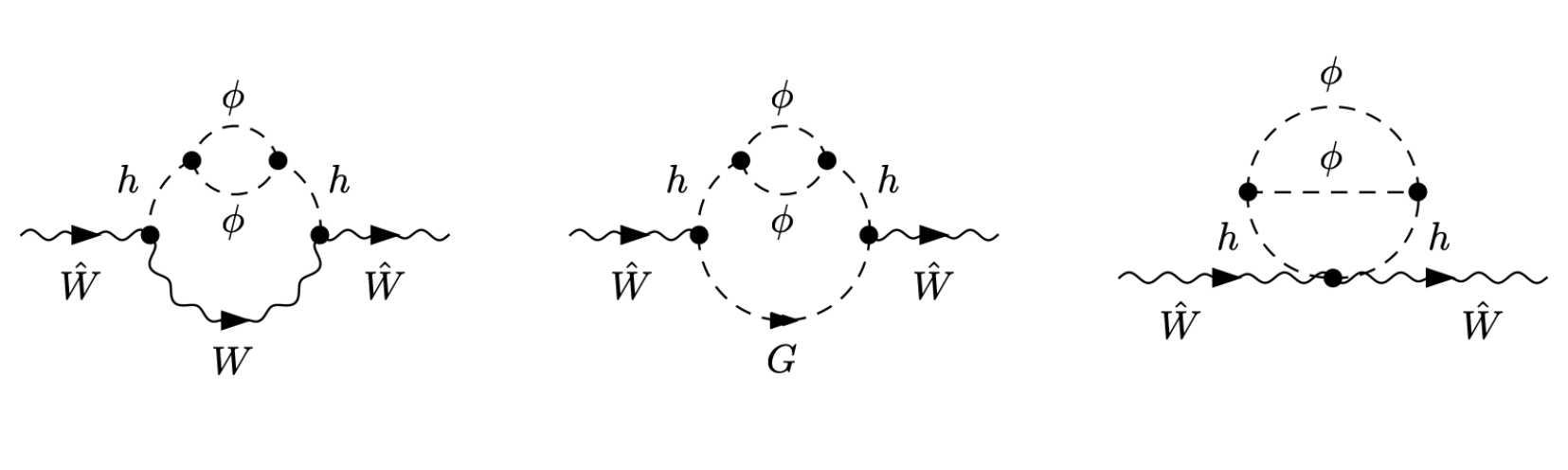}
\end{center}
\vspace{-10mm}
\caption{\label{fig:diagramsk2} Two-loop contributions to $\Pi_{WW}(q^2)$ proportional to $\kappa^2$, where $\kappa$ is the quartic portal coupling defined in~\eqref{eq:Lphi}. The field $\hat{W}$ denotes the background $W$-boson field in the background-field formalism, $W$ and $G$ are the quantum $W$ boson and the charged would-be Goldstone boson, and $h$ and $\phi$ denote the Higgs boson and the singlet scalar, respectively.}
\end{figure}

\begin{figure}[t!]
\begin{center}
\includegraphics[width=0.95\textwidth]{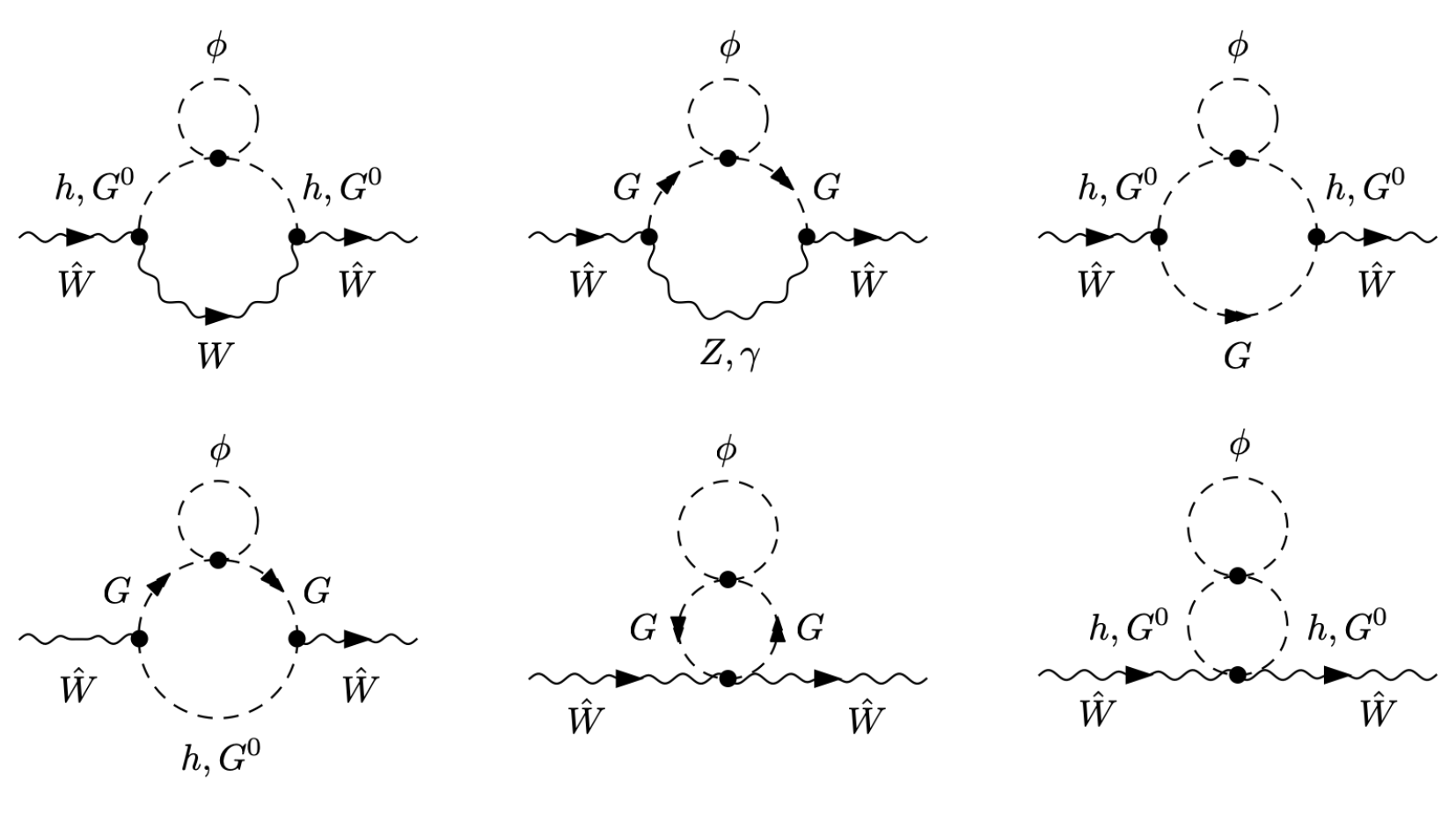}
\end{center}
\vspace{-10mm}
\caption{\label{fig:diagramsk1} Two-loop contributions to $\Pi_{WW}(q^2)$ proportional to $\kappa$ that cancel exactly with the ${\rm Re}\hspace{0.25mm}A_0(m_\phi^2)$ counterterms in~\eqref{eq:dmh2} and~\eqref{eq:dth}. In addition to the fields defined in the caption of~Figure~\ref{fig:diagramsk2}, the diagrams involve the quantum $Z$ boson, the neutral would-be Goldstone boson $G^0$, and the photon~$\gamma$.}
\end{figure}

In the singlet-scalar model~\eqref{eq:Lphi}, the correlators $\Pi_{VV^\prime}(q^2)$ first receive $\phi$-dependent contributions at two loops. These arise from genuine two-loop diagrams and from one-loop diagrams with insertions of the counterterms for the Higgs wave function, mass, and tadpole. The two-loop diagrams contributing to $\Pi_{WW}(q^2)$ are shown in~Figures~\ref{fig:diagramsk2} and~\ref{fig:diagramsk1}, which contain the terms proportional to $\kappa^2$ and $\kappa$, respectively. The external $W$-boson lines are treated as background fields, while all internal fields are quantum fields. In the on-shell~(OS) scheme, the relevant counterterms are~\cite{He:2016sqr,Englert:2019eyl,Haisch:2022rkm}
\begin{gather}
\delta Z_h = -\frac{\kappa^2 \hspace{0.125mm} v^2}{32 \hspace{0.125mm} \pi^2} \, {\rm Re} \hspace{0.25mm} B_0^\prime (m_h^2, m_\phi^2, m_\phi^2 ) \,, \label{eq:dZh} \\[2mm]
\delta m_h^2 = \frac{1}{32 \hspace{0.125mm} \pi^2} \, \Big [ \kappa^2 \hspace{0.125mm} v^2 \hspace{0.5mm} {\rm Re} \hspace{0.25mm} B_0 (m_h^2, m_\phi^2, m_\phi^2 ) + \kappa \hspace{0.5mm} {\rm Re} \hspace{0.25mm} A_0 (m_\phi^2 ) \Big ] \,, \label{eq:dmh2} \\[2mm]
\delta t_h = -\frac{\kappa \hspace{0.25mm} v}{32 \hspace{0.125mm} \pi^2} \, {\rm Re} \hspace{0.25mm} A_0 (m_\phi^2) \,, \label{eq:dth}
\end{gather}
where $v \simeq 246\,{\rm GeV}$ is the Higgs vacuum expectation value, $A_0(m^2)$, $B_0(q^2,m^2,m^2)$, and $B_0^\prime(q^2,m^2,m^2)$ are the standard one- and two-point Passarino-Veltman scalar integrals in the conventions of~\cite{Hahn:1998yk,Hahn:2016ebn}, and the prime again denotes a derivative with respect to $q^2$.

The calculation is performed in dimensional regularisation~(DR) with $d = 4 - 2\hspace{0.125mm}\epsilon$ and renormalisation scale~$\mu$, using the Feynman gauge of the background-field formalism~\cite{Abbott:1980hw,Abbott:1981ke,Denner:1994xt,Denner:1996gb}. The amplitudes are generated and evaluated with the \texttt{Mathematica} packages \texttt{FeynArts}~\cite{Hahn:2000kx}, \texttt{FeynCalc}~\cite{Shtabovenko:2020gxv}, \texttt{FormCalc}~\cite{Hahn:2016ebn}, and \texttt{Package-X}~\cite{Patel:2016fam}. The loop integrals are reduced to scalar master integrals with the Tarasov algorithm~\cite{Tarasov:1997kx}, as implemented in the \texttt{TARCER} package~\cite{Mertig:1998vk} bundled with \texttt{FeynCalc}, and the small-$q^2$ expansion~\eqref{eq:Piexp} is performed following~\cite{Davydychev:1992mt}. The resulting two-loop vacuum integrals are reduced to master integrals using the in-house \texttt{Mathematica} package \texttt{Coltrane}~\cite{Haisch:2002zz}, which implements partial-fraction decompositions, integration-by-parts~\cite{Tkachov:1981wb,Chetyrkin:1981qh}, and dimensional recurrence relations~\cite{Tarasov:1996br,Tarasov:1997kx} for one- and two-loop vacuum integrals with arbitrary masses. The same approach has previously been employed in~\cite{Haisch:2011up}.

The renormalised result has two notable structural features, both valid for arbitrary~$q^2$. First, the wave-function counterterm $\delta Z_h$ drops out identically: by the EW Ward identities, its contribution to the internal Higgs propagator is cancelled by its contribution to the adjacent vertices, independently of the gauge. Second, all terms linear in~$\kappa$ cancel between the bare two-loop diagrams and the counterterm contributions,
\beq
\hat \Pi_{VV^\prime}^{(2)}(q^2) \Big|_{\mathcal{O}(\kappa)} = \left[\Pi_{VV^\prime}^{(2)\hspace{0.25mm}\mathrm{bare}}(q^2) + \Pi_{VV^\prime}^{(2)\hspace{0.25mm}\mathrm{ct}}(q^2)\right]\Big|_{\mathcal{O}(\kappa)} = 0 \,.
\label{eq:kapcancel}
\eeq
For $\Pi_{WW}(q^2)$, the diagrams of Figure~\ref{fig:diagramsk1} are cancelled exactly by the ${\rm Re}\hspace{0.25mm}A_0(m_\phi^2)$ terms in~\eqref{eq:dmh2} and~\eqref{eq:dth}. This cancellation is required by decoupling: since ${\rm Re}\hspace{0.25mm}A_0(m_\phi^2) \propto m_\phi^2$, any leftover contribution of this form would grow quadratically with the singlet mass, in conflict with the Appelquist-Carazzone theorem~\cite{Appelquist:1974tg}. The renormalised correlators are therefore of~$\mathcal{O}(\kappa^2)$ and receive contributions only from diagrams of the type shown in~Figure~\ref{fig:diagramsk2} and from the ${\rm Re}\hspace{0.25mm}B_0(m_h^2, m_\phi^2, m_\phi^2)$ term in~\eqref{eq:dmh2}, whose role in the EFT interpretation of our results is discussed in~Section~\ref{sec:SMEFT}. Since the one-loop photon self-energies contain no internal Higgs line, $\hat \Pi_{\gamma\gamma}(q^2)$ and $\hat \Pi_{\gamma Z}(q^2)$ receive no such contributions.

\begin{figure}[t!]
\begin{center}
\includegraphics[width=0.6\textwidth]{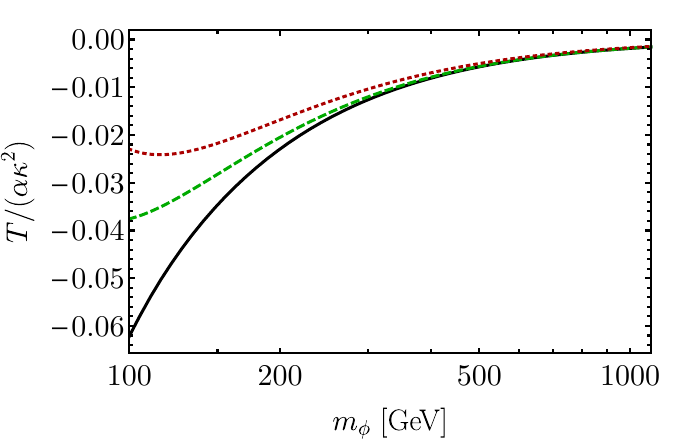} 

\vspace{0mm}

\includegraphics[width=0.6\textwidth]{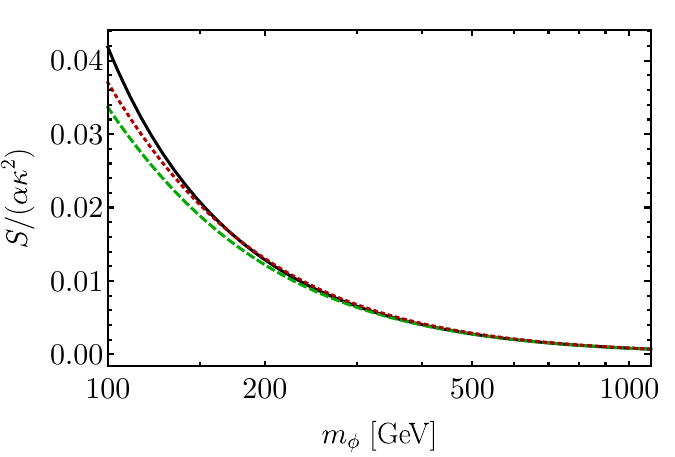}
\end{center}
\vspace{-4mm} 
\caption{\label{fig:TScomp} Comparison of the full two-loop results for $T$ (upper panel) and $S$ (lower panel) with the approximations given in~\eqref{eq:Tlimit} and~\eqref{eq:Slimit}, as a function of the singlet mass $m_\phi$. The solid black lines show the full results, while the dashed green and dotted red lines correspond to the approximations given in the second and third lines of~\eqref{eq:Tlimit} and~\eqref{eq:Slimit}, respectively. All results are normalised to $\alpha\hspace{0.25mm}\kappa^2$.}
\end{figure}

For $V = W, Z$ we~find
\begin{align}
\hat{\Pi}_{VV}^{(2)}(0) &= \frac{\kappa^2 \hspace{0.125mm} m_\phi^2}{256\hspace{0.25mm}\pi^4} \, \Bigg\{ 
\frac{x_V}{8} \left( \frac{1}{\epsilon} + 2 \ln x_\mu \right) 
- \frac{x_V\left(x_h + 47\hspace{0.25mm}x_V\right)}{16\left(x_h - x_V\right)} 
+ \frac{3\hspace{0.25mm}x_V^2}{2\left(x_h - x_V\right)} \ln x_V \nonumber \\[1mm]
& \hspace{-1cm} + \frac{3\hspace{0.25mm}x_V^2\left(x_h + x_V\right)}{2\left(x_h - x_V\right)^2} \, \ln \frac{x_h}{x_V} 
+ \frac{3\hspace{0.25mm}x_V^2\left[x_h\left(2 - x_V\right) + 2\hspace{0.25mm}x_V\right]}{4\hspace{0.25mm}x_h\left(x_h - x_V\right)^2}\,\phi\!\left(\frac{x_h}{4}\right) \nonumber \\[1mm]
& \hspace{-1cm} + \frac{3\hspace{0.25mm}\left(x_V - 4\right)\hspace{0.25mm}x_V^2}{4\left(x_h - x_V\right)^2}\,\phi\!\left(\frac{x_V}{4}\right) \nonumber \\[1mm]
& \hspace{-1cm} + \left[ \frac{x_V\left(x_h + 5\hspace{0.25mm}x_V\right)}{4\left(x_h - x_V\right)} 
- \frac{3\hspace{0.25mm}x_V^3}{2\left(x_h - x_V\right)^2} \, \ln \frac{x_h}{x_V} \right] \Big[ \mathrm{Re} \hspace{0.25mm} B(x_h) - \ln x_\mu \Big] \Bigg\} \,, \label{eq:PiVV0} \\[2mm]
\hat{\Pi}_{VV}^{\prime \hspace{0.125mm} (2)}(0) &= \frac{\kappa^2}{256\hspace{0.25mm}\pi^4} \, \Bigg\{ 
- \frac{x_V\left(2\hspace{0.25mm}x_h^2 - 3\hspace{0.25mm}x_hx_V + 61\hspace{0.25mm}x_V^2\right)}{6\left(x_h-x_V\right)^3} \nonumber \\[1mm]
& \hspace{-1cm} + \frac{x_V\left(x_h^3-4\hspace{0.25mm}x_h^2x_V+51\hspace{0.25mm}x_hx_V^2+12\hspace{0.25mm}x_V^3\right)}{6\left(x_h-x_V\right)^4} \, \ln \frac{x_h}{x_V} \nonumber \\[1mm]
& \hspace{-1cm} + \frac{x_V\left[x_h^2\left(x_V - 4\right) + x_h x_V\left(x_V + 6\right) + 2\hspace{0.25mm}x_V^2\left(14\hspace{0.25mm}x_V - 61\right)\right]}{6\left(x_h-x_V\right)^3\left(x_V-4\right)} \ln x_V \nonumber \\[1mm]
& \hspace{-1cm} + \frac{x_V\left[x_h^3 - x_h^2x_V\left(4+9\hspace{0.25mm}x_V\right) + 3\hspace{0.25mm}x_h x_V^2\left(17-2\hspace{0.25mm}x_V\right) + 12\hspace{0.25mm}x_V^3\right]}{6\hspace{0.25mm}x_h\left(x_h-x_V\right)^4}\,\phi\!\left(\frac{x_h}{4}\right) \nonumber \\[1mm]
& \hspace{-1cm} - \frac{x_V}{6\left(x_h-x_V\right)^4\left(x_V-4\right)} \, \Big[ x_h^2\left(6+x_V\right) - x_hx_V\left(76-50\hspace{0.25mm}x_V+9\hspace{0.25mm}x_V^2\right) \nonumber \\[1mm]
& \hspace{1cm} - x_V^2\left(170-69\hspace{0.25mm}x_V+6\hspace{0.25mm}x_V^2\right) \Big] \, \phi\!\left(\frac{x_V}{4}\right) \nonumber \\[1mm]
& \hspace{-1cm} + \left[ \frac{x_V\left(x_h^2+x_hx_V+28\hspace{0.25mm}x_V^2\right)}{6\left(x_h-x_V\right)^3}
- \frac{x_V^3\left(3\hspace{0.25mm}x_h+2\hspace{0.25mm}x_V\right)}{\left(x_h-x_V\right)^4} \, \ln \frac{x_h}{x_V} \right] \Big[ \mathrm{Re} \hspace{0.25mm} B(x_h) - \ln x_\mu \Big] \Bigg\} \,, \label{eq:PiVVprime0}
\end{align}
where $x_V = m_V^2/m_\phi^2$, $x_h = m_h^2/m_\phi^2$, and $x_\mu = \mu^2/m_\phi^2$. The loop functions~$B(z)$ and $\phi(z)$ are given in~Appendix~\ref{app:loopfun}. Since $\mathrm{Re}\hspace{0.25mm}B(x_h) - \ln x_\mu$ is $\mu$ independent, the derivative~\eqref{eq:PiVVprime0} is UV finite and scale independent. The self-energy~\eqref{eq:PiVV0} instead retains a residual $1/\epsilon$ pole and the associated logarithm $\ln x_\mu$, both proportional to~$x_V$. They drop out of $T$, because after division by $m_V^2$ they are the same for $V = W$ and $V = Z$, and they do not affect $S$, because they are independent of $q^2$.

While exact, the expressions~\eqref{eq:PiVV0} and~\eqref{eq:PiVVprime0} are rather lengthy. Their physics content becomes transparent in the decoupling limit $m_\phi \gg m_h, m_W, m_Z$. Expanding to first order in $x_h$, $x_W$, and $x_Z$, we obtain
\begin{align}
T^{(2)} & = \frac{1}{\alpha} \left [ \frac{\hat \Pi^{(2)}_{WW}(0)}{m_W^2} - \frac{\hat \Pi^{(2)}_{ZZ}(0)}{m_Z^2} \right ] \nonumber \\[1mm]
& \simeq \frac{\kappa^2\hspace{0.25mm} m_h^2}{256\hspace{0.125mm}\pi^4\hspace{0.25mm}\alpha\hspace{0.25mm}m_\phi^2} \left[ \frac{2}{3}\left(r_W - r_Z\right) - \frac{r_W^2 \ln r_W}{4\left(r_W - 1\right)} + \frac{r_Z^2 \ln r_Z}{4\left(r_Z - 1\right)} + \frac{r_W - r_Z}{4} \hspace{0.25mm} \ln \frac{m_\phi^2}{m_h^2} \right] \nonumber \\[1mm]
& \simeq -\frac{\kappa^2\hspace{0.25mm} v^2}{512\hspace{0.125mm}\pi^3\hspace{0.25mm}c_w^2\hspace{0.25mm}m_\phi^2} \left( \ln\frac{m_\phi}{m_h} + \frac{7}{12} \right) , \label{eq:Tlimit} \\[2mm]
S^{(2)} & = \frac{4 \hspace{0.125mm} s_w^2 \hspace{0.125mm} c_w^2}{\alpha} \, \hat \Pi^{\prime \hspace{0.125mm} (2)}_{ZZ}(0) \nonumber \\[1mm]
& \simeq \frac{s_w^2 \hspace{0.125mm} c_w^2 \hspace{0.25mm} \kappa^2\hspace{0.25mm} m_h^2}{64\hspace{0.125mm}\pi^4\hspace{0.25mm}\alpha\hspace{0.25mm}m_\phi^2} \left[ \frac{r_Z\left(13\hspace{0.25mm}r_Z^2 - 2\hspace{0.25mm}r_Z + 4\right)}{54\left(r_Z - 1\right)^2} - \frac{r_Z^3\left(r_Z + 9\right) \ln r_Z}{36\left(r_Z - 1\right)^3} + \frac{r_Z}{36} \hspace{0.25mm} \ln \frac{m_\phi^2}{m_h^2} \right] \nonumber \\[1mm]
& \simeq \frac{\kappa^2\hspace{0.25mm} v^2}{1152\hspace{0.125mm}\pi^3\hspace{0.25mm}m_\phi^2} \left( \ln\frac{m_\phi}{m_h} + \frac{23}{12} \right) , \label{eq:Slimit}
\end{align}
where $r_V = m_V^2/m_h^2$. The last lines of~\eqref{eq:Tlimit} and~\eqref{eq:Slimit} correspond to $m_h = m_Z$, and for~$T$ only the leading term in $s_w^2$ has been kept.

These approximations exhibit all the expected properties. The dependence on both~$1/\epsilon$ and $\mu$ cancels, so that $T^{(2)}$ and $S^{(2)}$ are UV finite and scale independent. Both parameters scale as $\kappa^2 v^2/m_\phi^2$ and hence decouple for $m_\phi \to \infty$. Moreover, $T^{(2)}$ vanishes for $r_W = r_Z$, as required for a measure of custodial-symmetry breaking. Finally, both parameters contain logarithms $\ln \hspace{0.125mm}(m_\phi/m_h)$ that are enhanced for a heavy singlet, reflecting the large hierarchy between the singlet mass and the EW scale. The origin of these logarithms is clarified in~Section~\ref{sec:IRlogs}.

The quality of the approximations is assessed in~Figure~\ref{fig:TScomp}, where they are compared with the full two-loop results for $T$ (upper panel) and $S$ (lower panel). For singlet masses $m_\phi \gtrsim 300\,{\rm GeV}$, both approximations reproduce the full results well. For lighter singlets, the approximation of $S$ remains accurate to about $20\%$, whereas the approximations of~$T$~underestimate the magnitude of the full result significantly, by up to a factor of two to three at $m_\phi = 100\,{\rm GeV}$. These deviations reflect the breakdown of the expansion in $m_h^2/m_\phi^2$. In EFT language, they signal that operators of dimension higher than six become numerically relevant once the singlet is not much heavier than the Higgs boson, with $T$ being considerably more sensitive to such effects than $S$.

A first two-loop calculation of $S$, $T$, and $U$ in this model was presented in~\cite{Englert:2020gcp}. While the counterterms and the cancellation of the terms linear in the portal coupling agree with our findings, the numerical results for the oblique parameters differ. Identifying the portal coupling of~\cite{Englert:2020gcp} as $\lambda = \kappa/2$, we obtain for $m_\phi = 100\,{\rm GeV}$ and $\lambda = 1$ the values $T^{(2)} \simeq -1.9 \cdot 10^{-3}$, $S^{(2)} \simeq 1.3 \cdot 10^{-3}$, and $U^{(2)} \simeq -3.5 \cdot 10^{-5}$, compared to $T \simeq -3.5 \cdot 10^{-4}$, $S \simeq -1.0 \cdot 10^{-3}$, and $U \simeq 6 \cdot 10^{-5}$ as read off from Figure~10 of~\cite{Englert:2020gcp}. Our signs of $T$ and~$S$ agree with the general argument presented in~Section~\ref{sec:IRlogs}, and our results are reproduced independently by the SMEFT calculation of~Section~\ref{sec:SMEFT} and by the dispersive representation of Appendix~\ref{app:dispersive}.

\section{IR logarithms from modified Higgs couplings}
\label{sec:IRlogs}

The logarithms $\ln \hspace{0.125mm}(m_\phi/m_h)$ found in~\eqref{eq:Tlimit} and~\eqref{eq:Slimit} have a simple physical origin, which we discuss in this section. A universal rescaling $g_{hVV} \to \kappa_V\hspace{0.125mm} g_{hVV}$ of the Higgs couplings to the EW gauge bosons, as parametrised in the $\kappa$-framework~\cite{LHCHiggsCrossSectionWorkingGroup:2012nn,LHCHiggsCrossSectionWorkingGroup:2013rie,LHCHiggsCrossSectionWorkingGroup:2016ypw}, spoils the cancellation of logarithmic UV divergences between the Higgs and would-be Goldstone boson contributions to the gauge-boson self-energies. The incomplete cancellation leaves a logarithmically enhanced shift in the oblique parameters~\cite{Barbieri:2007bh,Contino:2010rs,Panico:2015jxa},
\beq
\Delta T = -\frac{3}{16\hspace{0.125mm}\pi\hspace{0.125mm}c_w^2}\left(1-\kappa_V^2\right)\ln\frac{\Lambda^2}{m_h^2}\,, \qquad 
\Delta S = \frac{1}{12\hspace{0.125mm}\pi}\left(1-\kappa_V^2\right)\ln\frac{\Lambda^2}{m_h^2}\,, 
\label{eq:IRlogs}
\eeq
where $\Lambda$ denotes the scale at which the modification of the Higgs couplings is generated. From the EFT perspective, these logarithms describe the renormalisation group~(RG) evolution, from $\Lambda$ down to the EW scale, of the operators that define $T$ and $S$, driven by the operator that modifies~$\kappa_V$.

In the singlet-scalar model, $\kappa_V$ is modified by the operator~$Q_{H\Box}$. This is most easily seen by writing the Higgs doublet in polar form,
\beq
H = \frac{v+h}{\sqrt{2}} \, U(\pi) \begin{pmatrix} 0 \\ 1 \end{pmatrix} , \qquad U(\pi) = \exp \left( \frac{i\hspace{0.25mm}\pi^a \sigma^a}{v} \right) ,
\label{eq:polar}
\eeq
where the unitary matrix $U(\pi)$ contains the would-be Goldstone bosons $\pi^a$. Since $|H|^2 = (v+h)^2/2$ is independent of $\pi^a$, the operator~$Q_{H\Box}$ modifies only the kinetic term of the Higgs boson, while the Goldstone sector, and with it the masses and couplings of the longitudinal gauge bosons, remains SM-like. Canonically normalising the Higgs field then yields
\beq
\kappa_V \simeq 1 + v^2 \hspace{0.125mm} C_{H\Box} (m_\phi) \,,
\label{eq:kappaV}
\eeq
so that $1 - \kappa_V^2 \simeq -2\hspace{0.25mm}v^2\hspace{0.125mm} C_{H\Box} (m_\phi)$. Inserting the one-loop matching correction~\eqref{eq:high} into~\eqref{eq:IRlogs} and identifying $\Lambda = m_\phi$, we obtain
\beq
\Delta T^{(2)} \simeq -\frac{\kappa^2\hspace{0.25mm} v^2}{512\hspace{0.25mm}\pi^3\hspace{0.25mm}c_w^2\hspace{0.25mm}m_\phi^2} \ln\frac{m_\phi}{m_h} \,, \qquad 
\Delta S^{(2)} \simeq \frac{\kappa^2\hspace{0.25mm} v^2}{1152\hspace{0.25mm}\pi^3\hspace{0.25mm}m_\phi^2} \ln\frac{m_\phi}{m_h} \,.
\label{eq:IRlogssinglet}
\eeq
These expressions reproduce exactly the logarithmically enhanced terms of the full two-loop results~\eqref{eq:Tlimit} and~\eqref{eq:Slimit}, which constitutes a non-trivial cross-check of the calculation presented in~Section~\ref{sec:fulltwo}. At the same time, the comparison shows that~\eqref{eq:IRlogs} captures only the leading-logarithmic~(LL) part of the two-loop corrections, while the constant terms in~\eqref{eq:Tlimit} and~\eqref{eq:Slimit} lie beyond its reach. To account for them, a complete two-loop calculation in the SMEFT is required, to which we turn next.

\section{Oblique parameters in the SMEFT}
\label{sec:SMEFT}

The LL argument of the previous section can be put on a more solid footing by showing how the logarithms~\eqref{eq:IRlogs} emerge from a complete two-loop calculation in the SMEFT. Such a calculation involves two ingredients. The first is the one-loop contribution of the operator~$Q_{H\Box}$ to the gauge-boson self-energies, evaluated in the broken phase at the EW scale, which gives rise to the logarithms~\eqref{eq:IRlogs}. The second is the two-loop matching of the singlet-scalar model~\eqref{eq:Lphi} onto the SMEFT at a scale of order $m_\phi$, in particular onto the operators that contribute to $T$ and $S$ at tree level. For further details on two-loop calculations in the SMEFT, encompassing both matching and matrix-element computations, see, for instance,~\cite{Deutschmann:2017qum,Haisch:2022nwz,DiNoi:2023ygk,Gauld:2023gtb,Heinrich:2023rsd,Haisch:2024wnw,Haisch:2025lvd,DiNoi:2025uhu,Haisch:2025vqj,Haisch:2026eyq}.

\begin{figure}[t!]
\begin{center}
\includegraphics[width=0.95\textwidth]{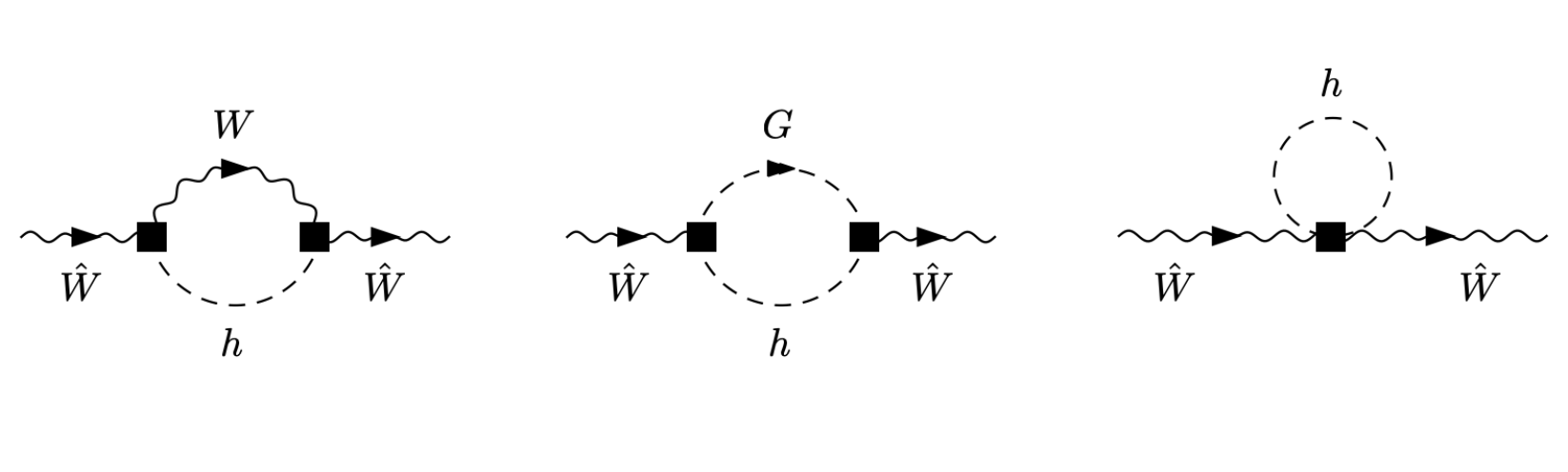}
\end{center}
\vspace{-10mm}
\caption{\label{fig:oneme} One-loop contributions to $\Pi_{WW}(q^2)$ arising from the operator~$Q_{H\Box}$. The fields are defined in the caption of Figure~\ref{fig:diagramsk2}. The black squares denote the Higgs-gauge boson vertices including the corrections induced by $Q_{H\Box}$. After expanding the product of the vertices, only the terms linear in $C_{H\Box}$ are retained, while the SM contributions and the quadratic terms are discarded.}
\end{figure}

We begin with the first ingredient. The one-loop diagrams that contribute to $\Pi_{WW}(q^2)$ at linear order in $C_{H\Box}$ are shown in~Figure~\ref{fig:oneme}, where the black squares represent the Higgs-gauge boson vertices including the $Q_{H\Box}$-induced corrections. The corresponding diagrams for $\Pi_{ZZ}(q^2)$ are obtained analogously. For $V = W, Z$, we~find the bare contributions
\begin{align}
\Delta \Pi_{VV}^{(2)}(0) &= -\frac{3\hspace{0.25mm}\alpha\hspace{0.25mm} v^2\hspace{0.25mm} m_V^4 \hspace{0.25mm} C_{H \Box}^{(1)}}{8\hspace{0.25mm}\pi\hspace{0.25mm} s_w^2\hspace{0.25mm} m_W^2} \left[ \frac{1}{\epsilon} + \frac{5\hspace{0.25mm} r_V - 1}{6\hspace{0.25mm} r_V} - \frac{r_V \ln r_V}{r_V - 1} + \ln r_\mu \right] , \label{eq:DeltaPiVV0} \\[2mm]
\Delta \Pi_{VV}^{\prime \hspace{0.125mm} (2)}(0) &= -\frac{\alpha\hspace{0.25mm} v^2\hspace{0.25mm} m_V^2 \hspace{0.25mm} C_{H \Box}^{(1)}}{24\hspace{0.25mm}\pi\hspace{0.25mm} s_w^2\hspace{0.25mm} m_W^2} \left[ \frac{1}{\epsilon} + \frac{41\hspace{0.25mm} r_V^2 + 14\hspace{0.25mm} r_V + 5}{6\left(r_V - 1\right)^2} - \frac{r_V^2\left(r_V + 9\right) \ln r_V}{\left(r_V - 1\right)^3} + \ln r_\mu \right] , \label{eq:DeltaPiVVprime0}
\end{align}
where $r_\mu = \mu^2/m_h^2$. The one-loop Wilson coefficient $C_{H\Box}^{(1)}$ is given in~\eqref{eq:high}. At the order considered here, its scale dependence can be neglected, and we identify $C_{H\Box}^{(1)} = C_{H\Box}^{(1)}(m_\phi)$. Since $C_{H\Box}$ is itself generated at one loop, the contributions~\eqref{eq:DeltaPiVV0} and~\eqref{eq:DeltaPiVVprime0} are formally of two-loop order.

The EFT results~\eqref{eq:DeltaPiVV0} and~\eqref{eq:DeltaPiVVprime0} are evaluated with the physical Higgs-boson mass, which matches the OS renormalisation employed in the full calculation. Indeed, as discussed in~Section~\ref{sec:fulltwo}, the only counterterm that survives there is the ${\rm Re}\hspace{0.25mm}B_0(m_h^2, m_\phi^2, m_\phi^2)$ term in~\eqref{eq:dmh2}, which subtracts the singlet-scalar contribution to the Higgs self-energy at $p^2 = m_h^2$. For $m_\phi \gg m_h$, the subtracted self-energy reduces to $-2\hspace{0.25mm}v^2\hspace{0.125mm} C_{H\Box}^{(1)} \left(p^2 - m_h^2\right)$, i.e.~to the modification of the Higgs kinetic term described by $Q_{H\Box}$. No scheme conversion between the full theory and the SMEFT is therefore required at the dimension-six level.

In~contrast to the full results~\eqref{eq:PiVV0} and~\eqref{eq:PiVVprime0}, the expressions~\eqref{eq:DeltaPiVV0} and~\eqref{eq:DeltaPiVVprime0} contain UV poles that do not cancel in $T$ and $S$. Inserting~\eqref{eq:DeltaPiVV0} and~\eqref{eq:DeltaPiVVprime0} into~\eqref{eq:defT} and~\eqref{eq:defS}, we~find
\beq
\Delta T^{(2)}_{H\Box} \Big|_{\rm pole} = \frac{3\hspace{0.25mm} v^2\hspace{0.25mm} C_{H\Box}^{(1)}}{8\hspace{0.25mm}\pi\hspace{0.25mm} c_w^2} \hspace{0.5mm} \frac{1}{\epsilon} \,, \qquad 
\Delta S^{(2)}_{H\Box} \Big|_{\rm pole} = -\frac{v^2\hspace{0.25mm} C_{H\Box}^{(1)}}{6\hspace{0.25mm}\pi} \hspace{0.5mm} \frac{1}{\epsilon} \,.
\label{eq:TSpoles}
\eeq
The origin of these poles is most transparent in the strongly-interacting light Higgs~(SILH) basis~\cite{Giudice:2007fh}, in which $T$ and $S$ are each described by a single Wilson coefficient or combination of Wilson coefficients. See, for instance,~\cite{Wells:2015uba} for a detailed discussion. The relevant operators~are
\beq
\begin{gathered}
O_H = \frac{1}{2}\hspace{0.25mm} \partial_\mu |H|^2 \hspace{0.25mm} \partial^\mu |H|^2 \,, \qquad 
O_T = \frac{1}{2} \hspace{0.25mm} \big(H^\dagger \overleftrightarrow{D}_\mu H\big)^2 \,, \\[1mm]
O_W = \frac{i g_L}{2} \hspace{0.25mm} \big(H^\dagger \sigma^a \overleftrightarrow{D}^\mu H\big) D^\nu W^a_{\mu\nu} \,, \qquad 
O_B = \frac{i g_Y}{2} \hspace{0.25mm} \big(H^\dagger \overleftrightarrow{D}^\mu H\big) \partial^\nu B_{\mu\nu} \,,
\end{gathered}
\label{eq:SILH}
\eeq
where $g_L$ and $g_Y$ denote the $SU(2)_L$ and $U(1)_Y$ gauge couplings, and, following~\cite{Giudice:2007fh}, $H^\dagger \overleftrightarrow{D}_\mu H = H^\dagger D_\mu H - (D_\mu H)^\dagger H$ and analogously for $H^\dagger \sigma^a \overleftrightarrow{D}_\mu H$. In the SILH basis, the covariant derivative acting on the Higgs doublet reads $D_\mu H = \big(\partial_\mu - i\hspace{0.25mm} g_L\hspace{0.25mm} \sigma^a W^a_\mu/2 - i\hspace{0.25mm} g_Y B_\mu/2\big) H$, which differs in the sign of the gauge couplings from the convention of~\cite{Grzadkowski:2010es}. At tree level, the oblique parameters are given in the SILH basis by
\beq
T = \frac{v^2}{\alpha}\hspace{0.5mm} c_T \,, \qquad 
S = \frac{4\hspace{0.125mm} s_w^2\hspace{0.25mm} m_W^2}{\alpha} \left( c_W + c_B \right) ,
\label{eq:TSSILH}
\eeq
while $O_H = -Q_{H\Box}/2$ up to a total derivative, so that $c_H = -2\hspace{0.25mm} C_{H\Box}$.

The poles~\eqref{eq:TSpoles} are cancelled by the counterterms of $c_T$ and $c_W + c_B$, which arise from the mixing of $O_H$ into $O_T$ and $O_W + O_B$. Writing $c_i^{(0)} = Z_{ij}\hspace{0.25mm} c_j$, where $c_i^{(0)}$ denote the bare Wilson coefficients and $Z_{ij} = \delta_{ij} + Z_{ij}^{(1)}/\epsilon$, the counterterm contributions read
\beq
\Delta T^{(2)}_{\mathrm{ct}} = -\frac{2\hspace{0.25mm} v^2}{\alpha} \hspace{0.5mm} Z_{TH}^{(1)} \hspace{0.5mm} \frac{C_{H\Box}^{(1)}}{\epsilon} \,, \qquad 
\Delta S^{(2)}_{\mathrm{ct}} = -\frac{8\hspace{0.25mm} s_w^2\hspace{0.25mm} m_W^2}{\alpha} \hspace{0.25mm} \Big( Z_{WH}^{(1)} + Z_{BH}^{(1)} \Big) \hspace{0.25mm} \frac{C_{H\Box}^{(1)}}{\epsilon} \,.
\label{eq:TSct}
\eeq
Requiring the sums of~\eqref{eq:TSpoles} and~\eqref{eq:TSct} to be finite yields
\beq
Z_{TH}^{(1)} = \frac{1}{16\hspace{0.25mm}\pi^2} \hspace{0.25mm} \frac{3\hspace{0.25mm} g_Y^2}{4} \,, \qquad 
Z_{WH}^{(1)} + Z_{BH}^{(1)} = -\frac{1}{16\hspace{0.25mm}\pi^2} \hspace{0.25mm} \frac{1}{3} \,,
\label{eq:ZSILH}
\eeq
where we have used $\alpha = g_Y^2 c_w^2/(4\pi)$ and $s_w^2 m_W^2/\alpha = \pi v^2$. Since $d c_i/d\ln\mu = 2\hspace{0.25mm} Z_{ij}^{(1)} c_j$ at one loop, the corresponding RG equations read
\beq
\frac{d c_T}{d\ln\mu} = \frac{1}{16\hspace{0.25mm}\pi^2} \frac{3\hspace{0.25mm} g_Y^2}{2}\hspace{0.5mm} c_H \,, \qquad 
\frac{d \big( c_W + c_B \big)}{d\ln\mu} = -\frac{1}{16\hspace{0.25mm}\pi^2} \hspace{0.25mm} \frac{2}{3} \hspace{0.5mm} c_H \,,
\label{eq:RGESILH}
\eeq
in agreement with~\cite{Elias-Miro:2013mua}. Solving these equations at LL accuracy between $\mu = m_\phi$ and the EW scale reproduces exactly the logarithms~\eqref{eq:IRlogssinglet}. The logarithmically enhanced terms in~$T$ and $S$ thus arise from the mixing of $O_H$, or equivalently $Q_{H\Box}$, into the operators that define the oblique parameters.

Removing the poles in the modified minimal subtraction~($\overline{\rm MS}$) scheme~\cite{Bardeen:1978yd} by adding the counterterms~\eqref{eq:TSct}, inserting the renormalised self-energies into~\eqref{eq:defT} and~\eqref{eq:defS}, setting $\mu = m_\phi$, and expanding to leading order in $v^2/m_\phi^2$, we obtain
\begin{align}
\Delta T_{H \Box}^{(2)} & = \frac{\kappa^2\hspace{0.25mm} m_h^2}{256\hspace{0.125mm}\pi^4\hspace{0.25mm}\alpha\hspace{0.25mm}m_\phi^2} \left[ \frac{5}{24}\left(r_W - r_Z\right) - \frac{r_W^2 \ln r_W}{4\left(r_W - 1\right)} + \frac{r_Z^2 \ln r_Z}{4\left(r_Z - 1\right)} + \frac{r_W - r_Z}{4} \hspace{0.25mm} \ln \frac{m_\phi^2}{m_h^2} \right] , \label{eq:TEFT} \\[2mm]
\Delta S_{H \Box}^{(2)} & = \frac{s_w^2 \hspace{0.125mm} c_w^2 \hspace{0.25mm} \kappa^2\hspace{0.25mm} m_h^2}{64\hspace{0.125mm}\pi^4\hspace{0.25mm}\alpha\hspace{0.25mm}m_\phi^2} \left[ \frac{r_Z\left(41\hspace{0.25mm}r_Z^2 + 14 \hspace{0.25mm}r_Z + 5\right)}{216\left(r_Z - 1\right)^2} - \frac{r_Z^3\left(r_Z + 9\right) \ln r_Z}{36\left(r_Z - 1\right)^3} + \frac{r_Z}{36} \hspace{0.25mm} \ln \frac{m_\phi^2}{m_h^2} \right] . \label{eq:SEFT}
\end{align}
These expressions reproduce not only the logarithms~\eqref{eq:IRlogssinglet}, but the complete non-linear dependence of~\eqref{eq:Tlimit} and~\eqref{eq:Slimit} on the ratios $r_W$ and $r_Z$. The expanded two-loop results~\eqref{eq:Tlimit} and~\eqref{eq:Slimit} can therefore be written as
\begin{align}
T^{(2)} & \simeq \Delta T_{H \Box}^{(2)} + \frac{\kappa^2\hspace{0.25mm} m_h^2}{256\hspace{0.125mm}\pi^4\hspace{0.25mm}\alpha\hspace{0.25mm}m_\phi^2} \left[ \frac{11}{24}\left(r_W - r_Z\right) \right] \simeq \Delta T_{H \Box}^{(2)} - \frac{\kappa^2\hspace{0.25mm} v^2}{512\hspace{0.125mm}\pi^3\hspace{0.25mm}c_w^2\hspace{0.25mm}m_\phi^2} \hspace{0.25mm} \frac{11}{12} \,, \label{eq:fullTEFT} \\[2mm]
S^{(2)} & \simeq \Delta S_{H \Box}^{(2)} + \frac{s_w^2 \hspace{0.125mm} c_w^2 \hspace{0.25mm} \kappa^2\hspace{0.25mm} m_h^2}{64\hspace{0.125mm}\pi^4\hspace{0.25mm}\alpha\hspace{0.25mm}m_\phi^2} \left[ \frac{11 \hspace{0.25mm} r_Z}{216} \right] \simeq \Delta S_{H \Box}^{(2)} + \frac{\kappa^2\hspace{0.25mm} v^2}{1152\hspace{0.125mm}\pi^3\hspace{0.25mm}m_\phi^2} \hspace{0.25mm} \frac{11}{12} \,. \label{eq:fullSEFT}
\end{align}
The remaining terms are linear in $r_W - r_Z$ and $r_Z$, respectively. Since $m_h^2\hspace{0.25mm}(r_W - r_Z) = -s_w^2\hspace{0.25mm} m_Z^2$ and $m_h^2\hspace{0.25mm} r_Z = m_Z^2$ are both proportional to $v^2$, they correspond to pure constants in units of $\kappa^2 v^2/m_\phi^2$, which are not reproduced by the EFT matrix elements of $Q_{H\Box}$. Instead, they correspond to the second ingredient of the SMEFT calculation, namely the two-loop matching corrections to $c_T$ and $c_W + c_B$ at $\mu = m_\phi$.

\begin{figure}[t!]
\begin{center}
\includegraphics[width=0.7\textwidth]{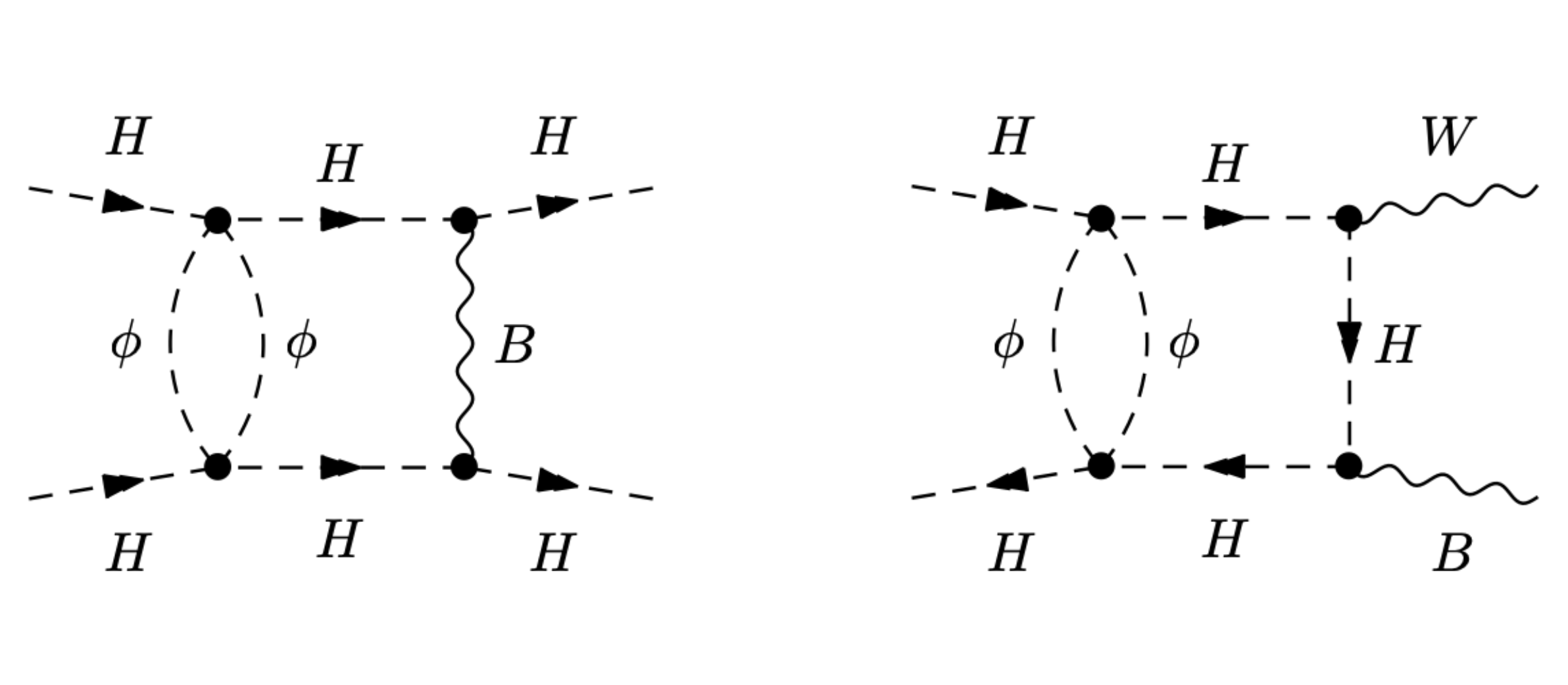}
\end{center}
\vspace{-10mm}
\caption{\label{fig:SMEFTdiagrams} Examples of two-loop diagrams contributing to the matching of the singlet-scalar model onto the SMEFT in the unbroken phase. Left: four-point function $H H \to H H$, which determines~$c_T$. Right: four-point function $H H^\dagger \to W B$, which determines~$c_W + c_B$. In both diagrams, the gauge bosons couple to the internal Higgs line between the two portal vertices. The~remaining diagrams follow from different attachments of the gauge bosons. See the main text for further~details.}
\end{figure}

To compute these matching corrections, we evaluate two-loop four-point functions in the unbroken phase, examples of which are shown in~Figure~\ref{fig:SMEFTdiagrams}. The Wilson coefficient~$c_T$ is extracted from the four-point function $H H \to H H$ displayed on the left-hand side, while the combination~$c_W + c_B$ follows from the four-point function $H H^\dagger \to W B$ shown on the right-hand side. In both cases, the gauge bosons couple to the internal Higgs line between the two portal vertices, which are connected by the singlet loop. All loop momenta are of the order of $m_\phi$, so that the masses of the light fields can be set to zero and the integrands can be expanded in the external momenta, reducing the calculation to two-loop vacuum integrals. The details of this calculation are given in~Appendix~\ref{sec:SMEFTdetails}. The resulting expressions contain IR poles that cancel exactly against the UV poles~\eqref{eq:TSpoles} of the EFT matrix elements, which provides an independent check of the anomalous dimensions~\eqref{eq:ZSILH}. After removing these poles, we obtain in the $\overline{\rm MS}$ scheme
\beq
c_T^{(2)} (m_\phi) = -\frac{1}{(16 \hspace{0.125mm} \pi^2)^2} \hspace{0.125mm} \frac{11 \hspace{0.25mm} g_Y^2 \hspace{0.25mm} \kappa^2}{96 \hspace{0.25mm} m_\phi^2}\,, \qquad
c_W^{(2)} (m_\phi) + c_B^{(2)} (m_\phi) = \frac{1}{(16 \hspace{0.125mm} \pi^2)^2} \hspace{0.125mm} \frac{11 \hspace{0.25mm} \kappa^2}{216 \hspace{0.25mm} m_\phi^2}\,.
\label{eq:cTcWBmatch}
\eeq
Inserting these results into~\eqref{eq:TSSILH} reproduces exactly the constant terms in~\eqref{eq:fullTEFT} and~\eqref{eq:fullSEFT}. Combined with the EFT matrix elements~\eqref{eq:TEFT} and~\eqref{eq:SEFT}, the SMEFT calculation therefore reproduces the full two-loop results~\eqref{eq:Tlimit} and~\eqref{eq:Slimit} completely at the dimension-six level, including the logarithmically enhanced terms, the dependence on the EW gauge-boson masses, and the constants. With the exact two-loop calculation, the IR logarithms from modified Higgs couplings, and the SMEFT description in full agreement, the triple-threat challenge for $T$ and $S$ at two loops is thus successfully met.

\section{Constraints from EWPO}
\label{sec:constraints}
We now turn to the phenomenological implications of our results. Among the EWPO, the $W$-boson mass and the effective weak mixing angle $s_{\rm eff}^2$ are particularly sensitive to the oblique parameters. We work in the LEP input scheme~\cite{ALEPH:2005ab,Brivio:2017vri,Biekotter:2023xle,Biekotter:2023vbh}, which takes $\alpha$, the Fermi constant $G_F$ extracted from muon decay, and the on-shell $Z$-boson mass $m_Z$ as input parameters. In this scheme, both the weak mixing angle and $m_W$ are derived quantities, with $m_W = c_w\hspace{0.25mm}m_Z$ at tree level. Neglecting the $U$ parameter, which is numerically irrelevant in our case as shown in~Appendix~\ref{app:UWY}, the relative shift of $m_W$ induced by new physics is given by~\cite{Peskin:1991sw}
\beq
\frac{\delta m_W}{m_W} \simeq \frac{\alpha}{2\left(c_w^2 - s_w^2\right)} \left( c_w^2\hspace{0.25mm} T - \frac{S}{2} \right) .
\label{eq:dmWmW}
\eeq
Similarly, the effective weak mixing angle, which is accessed at the $Z$ pole through the left-right asymmetry $A_{\rm LR}^0 = A_e$ at SLD and the forward-backward asymmetries $A_{\rm FB}^{0,f} = 3/4 \hspace{0.25mm} A_e A_f$, where $A_f = 2\hspace{0.25mm} g_V^f g_A^f/\big((g_V^f)^2 + (g_A^f)^2\big)$ is the asymmetry parameter of the $Z$-boson coupling to the fermion~$f$, with vector and axial-vector couplings $g_V^f = T_3^f - 2\hspace{0.25mm} Q_f\hspace{0.25mm} s_{\rm eff}^{2,f}$ and $g_A^f = T_3^f$. Here $T_3^f$ denotes the third component of the weak isospin of the left-handed fermion and~$Q_f$ its electric charge. For leptons, $T_3^\ell = -1/2$ and $Q_\ell = -1$, so that $A_\ell \simeq 2\hspace{0.25mm}(1 - 4\hspace{0.25mm} s_{\rm eff}^2)$, receives the relative shift
\beq
\frac{\delta s_{\rm eff}^2}{s_{\rm eff}^2} \simeq \frac{\alpha}{s_w^2 \left( c_w^2 - s_w^2 \right)} \left( \frac{S}{4} - s_w^2\hspace{0.25mm} c_w^2\hspace{0.25mm} T \right) .
\label{eq:ds2}
\eeq
In our numerical analysis, we employ $\alpha^{-1}(m_Z) = 128.952$, $G_F = 1.1663788 \cdot 10^{-5}\,{\rm GeV}^{-2}$, and $m_Z = 91.1879\,{\rm GeV}$, which in the LEP scheme imply $v = 246.22\,{\rm GeV}$ and $s_w^2 = 0.2310$, as well as $m_h = 125.20\,{\rm GeV}$~\cite{ParticleDataGroup:2026mpi}.

Since hadron colliders currently provide the most precise determinations of $m_W$, we adopt the preliminary combination of recent measurements by ATLAS, CMS, D\O, and LHCb~\cite{Bozzi:2025xtr} together with the SM prediction of~\cite{Awramik:2003rn},
\beq
m_W^{\rm exp} = \left ( 80.361 \pm 0.008 \right ) {\rm GeV} \,, \qquad m_W^{\rm SM} = \left ( 80.356 \pm 0.004 \right ) {\rm GeV} \,.
\label{eq:mwpresent}
\eeq
The two values agree within $0.6\hspace{0.25mm}\sigma$. Adding the uncertainties in quadrature, the relative shift is constrained at 95\%~confidence level~(CL) to
\beq
\frac{\delta m_W}{m_W} \in \left[ -1.56,\, 2.80 \right] \cdot 10^{-4} \,.
\label{eq:dmWmWnow}
\eeq
For the FCC-ee projections, we use the three theory scenarios of~\cite{Greljo:2025ggc}. Scenario~S1 (conservative theory progress) includes uncertainties on both the SM predictions and the extraction of the pseudo-observables from data, S2 (aggressive theory progress) retains only those on the SM predictions, and S3 keeps experimental errors only. The $WW$-threshold and Tera-$Z$ runs of FCC-ee are projected to shrink the interval~\eqref{eq:dmWmWnow} by more than an order of magnitude~\cite{Greljo:2025ggc,Haisch:2026eyq}, to $\pm\hspace{0.25mm} 9.7 \cdot 10^{-6}$, $\pm\hspace{0.25mm} 6.4 \cdot 10^{-6}$, and $\pm\hspace{0.25mm} 5.9 \cdot 10^{-6}$ in S1, S2, and S3, respectively. For $s_{\rm eff}^2$, we combine all asymmetries, treating the common theory error in S2 as correlated among them. This~gives $\delta s_{\rm eff}^2/s_{\rm eff}^2 \in \left[ -1.57,\, 1.26 \right] \cdot 10^{-3}$ for the LEP and SLD data and $\pm\hspace{0.25mm} 4.2 \cdot 10^{-5}$, $\pm\hspace{0.25mm} 7.2 \cdot 10^{-6}$, and $\pm\hspace{0.25mm} 5.5 \cdot 10^{-6}$ for FCC-ee in S1, S2, and S3, respectively.

\begin{figure}[t!]
\centering
\includegraphics[width=0.48\textwidth]{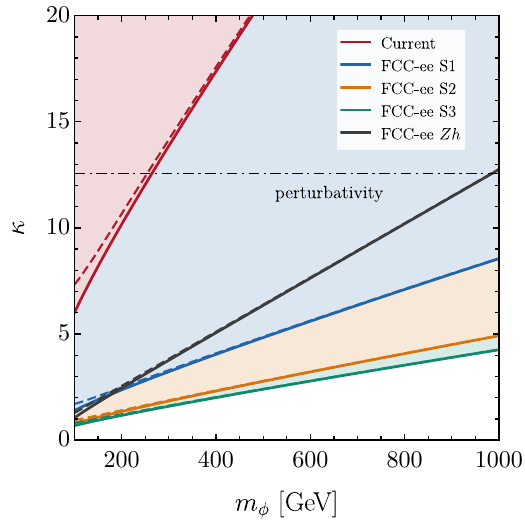}
\hfill
\includegraphics[width=0.48\textwidth]{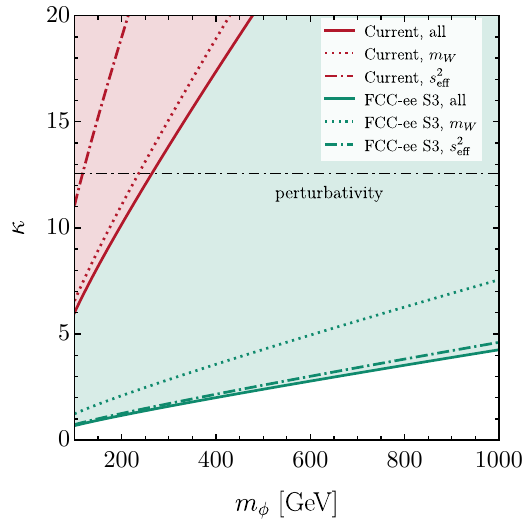}
\caption{95\%~CL exclusion regions in the $m_\phi\hspace{0.25mm}$--$\hspace{0.25mm}\kappa$ plane of the singlet-scalar model. Left panel: All~EWPO for current data and projections for FCC-ee under different assumptions about theoretical uncertainties~(S1, S2, S3). Solid lines employ the full two-loop results for $T$ and $S$, and dashed lines the dimension-six approximations in the second lines of~\eqref{eq:Tlimit} and~\eqref{eq:Slimit}. The black lines show the FCC-ee reach of $e^+e^- \to Zh$ with experimental errors only, using $\delta \sigma_{Zh}/\sigma_{Zh} = \delta Z_h$ with~\eqref{eq:dZh}~(solid) and its heavy-mass limit~\eqref{eq:Zh}~(dashed). Right panel: Two-loop bounds for current data and S3, from all observables~(solid), $m_W$ alone~(dotted), and $s_{\rm eff}^2$ alone~(dash-dotted). In both panels, the dash-dotted black line indicates the naive perturbativity bound $\kappa = 4\pi$. See the main text for additional details.}
  \label{fig:mvskcomp}
\end{figure}

Inserting the two-loop results for $T$ and $S$ into the predictions for all EWPO yields the 95\%~CL exclusion regions in the left panel of~Figure~\ref{fig:mvskcomp}. Since $T^{(2)} < 0$ and $S^{(2)} > 0$, both parameters decrease $m_W$ and increase $s_{\rm eff}^2$, so that their effects add up. For heavy singlets, the exclusion boundaries scale approximately linearly, $\kappa \propto m_\phi$, since both $T^{(2)}$ and $S^{(2)}$ scale as $\kappa^2 v^2/m_\phi^2$, up to a mild logarithmic dependence on $m_\phi$. Current data exclude $\kappa \gtrsim 6$ at $m_\phi = 100\,{\rm GeV}$ and $\kappa \gtrsim 14$ at $m_\phi = 300\,{\rm GeV}$. At $m_\phi = 100\,{\rm GeV}$, $300\,{\rm GeV}$, and $1\,{\rm TeV}$, FCC-ee excludes $\kappa \gtrsim 1.4$, $3.2$, and $8.6$ in S1, $\kappa \gtrsim 0.8$, $1.8$, and $4.9$ in S2, and $\kappa \gtrsim 0.7$, $1.6$, and $4.3$ in~S3.

The right panel shows that these bounds are essentially set by two observables. Since $A_e \simeq 8\hspace{0.25mm}(1/4 - s_{\rm eff}^2)$ is accidentally small, the asymmetries respond strongly to a shift in~$s_{\rm eff}^2$, with $\delta A_e/A_e \simeq -12\,\delta s_{\rm eff}^2/s_{\rm eff}^2$, whereas $\delta m_W/m_W \simeq -0.5\,\delta s_{\rm eff}^2/s_{\rm eff}^2$ for a shift in $T$ and $\delta m_W/m_W \simeq -s_w^2\,\delta s_{\rm eff}^2/s_{\rm eff}^2$ for a shift in $S$. Current bounds are dominated by $m_W$, since the present asymmetry measurements are not precise enough to compensate for their larger sensitivity. The same is true in S1, where theory errors limit the precision, while $s_{\rm eff}^2$ leads in~S2 and S3. Together, these two observables reproduce the full bounds to within about~6\%, with most of the remainder coming from the total $Z$-boson width~$\Gamma_Z$.

Figure~\ref{fig:mvskcomp} also allows us to assess the quality of the SMEFT description. The dashed lines retain only the leading terms in $v^2/m_\phi^2$ and thus correspond to the complete SMEFT result at the dimension-six level. For heavy singlets, the full and the SMEFT calculations give essentially identical bounds. For light singlets, however, the full calculation yields stronger constraints. This difference stems mainly from $T$, whose magnitude is underestimated by the dimension-six approximation at small $m_\phi$, as discussed in~Section~\ref{sec:fulltwo}. The relative shift of the bounds is approximately the same for all curves, about $20\%$ at $m_\phi = 100\,{\rm GeV}$ and~$2\%$~at $m_\phi = 300\,{\rm GeV}$. A dimension-six SMEFT estimate of the FCC-ee reach, without the two-loop matching constants, was given in~\cite{Maura:2024zxz}, with RG running from a fixed scale of $500\,{\rm GeV}$ rather than from~$m_\phi$. Rescaled to 95\%~CL and with its Lagrangian mass interpreted as the physical mass~$m_\phi$, a difference of dimension eight, its $Q_{H\Box}$-only bound agrees with our~S1~result to within about $15\%$ for $m_\phi \gtrsim 300\,{\rm GeV}$, as both are dominated by~$m_W$ with a similar projected uncertainty.

The derived limits should be compared with the requirement that the portal coupling remains perturbative, which roughly implies $\kappa \lesssim 4\pi$, as indicated by the dash-dotted black line in~Figure~\ref{fig:mvskcomp}. Current EWPO probe the perturbative regime only for light singlets with $m_\phi \lesssim 260\,{\rm GeV}$, above which the excluded couplings exceed $4\pi$. In~contrast, the FCC-ee EWPO projections remain below $4\pi$ over the entire mass range considered. Future EWPO measurements at FCC-ee would therefore provide sensitivity to perturbative Higgs portals with singlet masses up to the TeV scale.

It is also instructive to compare these bounds with the projected reach of collider probes of the Higgs portal. The high-luminosity LHC~(HL-LHC) is expected to exclude~$\kappa \gtrsim 3$ for $m_\phi \simeq 100\,{\rm GeV}$ and $\kappa \gtrsim 8$ for $m_\phi \simeq 250\,{\rm GeV}$~\cite{Ruhdorfer:2019utl,Haisch:2022rkm}. These projections combine searches for off-shell Higgs production in VBF and in $pp \to ZZ \to 4\ell$ with measurements of the Higgs signal strengths and of double-Higgs production. At the hadron-hadron option of the~FCC~(FCC-hh), VBF production of a pair of singlets probes $\kappa \simeq 3.5$ up to $m_\phi \simeq 330\,{\rm GeV}$~\cite{Ruhdorfer:2019utl}. The~current EWPO constraint is thus not competitive with the HL-LHC, while EWPO measurements at FCC-ee would reach a sensitivity comparable to that of FCC-hh. A particularly close connection exists with the measurement of the $e^+ e^- \to Zh$ cross section at FCC-ee. At one loop, the singlet modifies this cross section by~\cite{Craig:2013xia,Ruhdorfer:2019utl}
\beq
\frac{\delta \sigma_{Zh}}{\sigma_{Zh}} \simeq \kappa_V^2 - 1 \simeq -\frac{\kappa^2 v^2}{192 \pi^2 m_\phi^2} \,,
\label{eq:Zh}
\eeq
which, up to a sign, is the combination that multiplies the IR logarithms in~\eqref{eq:IRlogs}. Since at this order the singlet enters only through the Higgs wave function, the full mass dependence is obtained by replacing the right-hand side of~\eqref{eq:Zh} with $\delta Z_h$ in~\eqref{eq:dZh}.

Combining the $240\,{\rm GeV}$ and $365\,{\rm GeV}$ runs, with experimental errors of $0.31\%$ and $0.52\%$~\cite{FCC:2025lpp,Selvaggi:2025kmd}, excludes $\kappa \gtrsim 3.8$ for $m_\phi = 300\,{\rm GeV}$ and $\kappa \gtrsim 13$ for $m_\phi = 1\,{\rm TeV}$ at 95\%~CL, as shown in the left panel of~Figure~\ref{fig:mvskcomp}. This is of the same order as the FCC-ee EWPO in S1 below roughly $200\,{\rm GeV}$, and increasingly weaker above. The cross section and the EWPO thus provide complementary probes of the same Wilson coefficient~$C_{H\Box}$. Both are sensitive to part of the parameter space in which the singlet can induce a strong first-order EW phase transition, which extends over $m_\phi \simeq 150\,{\rm GeV}$~to~$m_\phi \simeq 550\,{\rm GeV}$ and $\kappa \simeq 2$ to $\kappa  \simeq 8$~\cite{Curtin:2014jma}, in our normalisation of $\kappa$.

\section{Conclusions}
\label{sec:conclusions}

In this article, we have studied the leading contributions to the oblique parameters $T$~and~$S$ in the~$Z_2$-symmetric real singlet-scalar extension of the SM. Because the $Z_2$ symmetry forbids mixing between the singlet and the Higgs boson, these contributions first arise at two loops. We have approached their calculation from three complementary angles, which together provide stringent cross-checks and a transparent physical picture.

First, we have performed an exact two-loop calculation of the gauge-boson self-energies in the full theory and derived compact expressions for a heavy singlet, which are UV finite, scale independent, and decouple as $\kappa^2 v^2/m_\phi^2$, but contain logarithms $\ln \hspace{0.125mm}(m_\phi/m_h)$. Second, we have shown that these logarithms arise because the singlet rescales the Higgs couplings to the EW gauge bosons, but not those of the would-be Goldstone bosons, thereby spoiling the cancellation of logarithmic UV divergences in the gauge-boson self-energies. Third, we have reproduced our results in the SMEFT, where the one-loop matrix elements of the operator that modifies the Higgs kinetic term account for the logarithms and all terms that depend on the EW masses. Their UV poles are removed by the mixing of this operator into the operators that define $T$ and $S$, while the remaining constants correspond to two-loop matching corrections to the latter. By performing the two-loop matching explicitly in the unbroken phase, we have verified these constants independently, thereby completing the SMEFT description of our results, and together these calculations establish results that differ from those of the earlier two-loop calculation~\cite{Englert:2020gcp}. Unlike $T$ and $S$, the two-loop corrections to $U$ contain no logarithms of $m_\phi/m_h$, and they are one to two orders of magnitude smaller. As shown in~Appendix~\ref{app:UWY}, the same holds for the parameters $W$ and $Y$ introduced in~\cite{Barbieri:2004qk}. Finally, in~Appendix~\ref{app:dispersive} we have derived a dispersive representation of the gauge-boson self-energies that is valid for arbitrary external momentum. It expresses the full two-loop results through one-loop SM functions integrated over the spectral density of the singlet pair, and so provides a further independent check of our calculation.

Comparing the full two-loop results with their leading terms in $v^2/m_\phi^2$, we find that the dimension-six approximation works well for $m_\phi \gtrsim 300\,{\rm GeV}$, whereas for lighter singlets operators of higher dimension become numerically relevant, in particular for $T$. These features carry over to the constraints from EWPO, where the full two-loop results lead to noticeably stronger bounds at low singlet masses. Current data exclude $\kappa \gtrsim 6$ at $m_\phi = 100\,{\rm GeV}$ and $\kappa \gtrsim 14$ at $m_\phi = 300\,{\rm GeV}$, and thus probe perturbative portal couplings only for $m_\phi \lesssim 260\,{\rm GeV}$. The projected FCC-ee sensitivity improves these limits considerably, excluding $\kappa \gtrsim 1.4$ at $m_\phi = 100\,{\rm GeV}$ and $\kappa \gtrsim 8.6$ at $m_\phi = 1\,{\rm TeV}$ in S1, and $\kappa \gtrsim 0.7$ and $\kappa \gtrsim 4.3$ in S3. FCC-ee would therefore probe perturbative Higgs portals with singlet masses up to the TeV scale, with a sensitivity comparable to that of FCC-hh. These measurements would also complement a precision measurement of the $e^+e^- \to Zh$ cross section, which probes the same Wilson coefficient~$C_{H\Box}$ that governs the~IR logarithms in $T$ and $S$.

Our analysis generalises directly to other new-physics sectors of SM gauge singlets that couple to the SM exclusively through $|H|^2$ and do not mix with the Higgs boson, such as multi-singlet portals or hidden sectors charged under an unbroken dark gauge group. In~these cases, only the Wilson coefficient of the operator that modifies the Higgs kinetic term and the two-loop matching constants change. As shown in~Appendix~\ref{app:dispersive}, at second order in the portal interaction, the two-loop corrections to all oblique parameters of such sectors are in fact fully determined by the spectral function of the hidden-sector operator that couples to $|H|^2$. The IR logarithms also determine the logarithmically enhanced two-loop corrections to $T$ in other extensions of the Higgs sector with custodially symmetric one-loop effects, such as the custodial quadruplet model~\cite{Durieux:2022hbu,Maura:2024zxz,Maura:2025rcv} or the two-Higgs-doublet model in the alignment limit with degenerate CP-odd and charged Higgs bosons~\cite{Bahl:2022jnx,Stylianou:2023xit}. Both~allow for sizeable modifications of the Higgs self-couplings, while the corresponding two-loop matching constants require a model-specific calculation, which we leave for future~work.

A further natural extension of our study is the combination of our results with measurements of the Higgs self-couplings, which would offer a more complete picture of Higgs-portal physics at future colliders.

\acknowledgments
The Feynman diagrams in this article were drawn with \texttt{FeynArts}~\cite{Hahn:2000kx}, while the plots were produced with the help of \texttt{MaTeX}~\cite{MaTeX}. During the preparation of this work, the authors used artificial intelligence~(AI) in the form of the large language models Claude~Opus~5.5~(Anthropic) and GPT-6 Astra~(OpenAI). It assisted with improving the language and \LaTeX{} typesetting of the manuscript and with cross-checking selected analytic manipulations. We can assure the reader that no superintelligence~(SI) was involved at any stage, the only SI entering this work being the International System of Units, which we have nevertheless managed to avoid by quoting all results in natural units. The authors reviewed and edited all AI-assisted content and remain fully responsible for all results presented in this article, including any remaining mistakes, which are therefore of entirely natural origin. SM thanks the International Max~Planck Research School (IMPRS) on “Elementary Particle Physics” as well as the Collaborative~Research Center SFB1258 for support.

\begin{appendix}

\section{Loop functions}
\label{app:loopfun}

This appendix collects the loop functions entering the two-loop gauge-boson self-energies of Section~\ref{sec:fulltwo}. The function $B(z)$ is the UV-finite part of the Passarino-Veltman two-point scalar integral with two equal internal masses in the conventions of~\cite{Hahn:1998yk,Hahn:2016ebn}. It reads
\beq \label{eq:B}
B(z) = 2 + \ln x_\mu + \frac{\sqrt{z(z-4)}}{z}\ln\frac{2 - z + \sqrt{z(z-4)}}{2} \,,
\eeq
with $x_\mu = \mu^2/m_\phi^2$ as defined in Section~\ref{sec:fulltwo}.

The function $\phi(z)$ arises from the two-loop vacuum integral with two different masses, evaluated in closed form in~\cite{Davydychev:1992mt}. It is given by
\beq \label{eq:phi}
\phi(z) = 
\begin{cases}
4\sqrt{\dfrac{z}{1-z}}\,\operatorname{Cl}_2\!\left(2\arcsin\sqrt{z}\right) \,, & z < 1 \,, \\[4mm]
\dfrac{1}{\lambda}\left[-4\operatorname{Li}_2\!\left(\dfrac{1-\lambda}{2}\right) + 2\ln^2\!\left(\dfrac{1-\lambda}{2}\right) - \ln^2(4z) + \dfrac{\pi^2}{3}\right] , & z > 1 \,,
\end{cases}
\eeq
where we introduced $\lambda = \sqrt{1 - 1/z}$, $\operatorname{Cl}_2(z) = \operatorname{Im}\hspace{0.125mm}[\operatorname{Li}_2\hspace{0.125mm}(e^{iz})]$ is the Clausen function, and $\operatorname{Li}_2(z) = -\int_0^z dt\,\ln(1-t)/t$ is the dilogarithm.

\section{The $\bm{U}$, $\bm{W}$, and $\bm{Y}$ parameters}
\label{app:UWY}

\begin{figure}[t!]
\begin{center}
\includegraphics[width=0.6\textwidth]{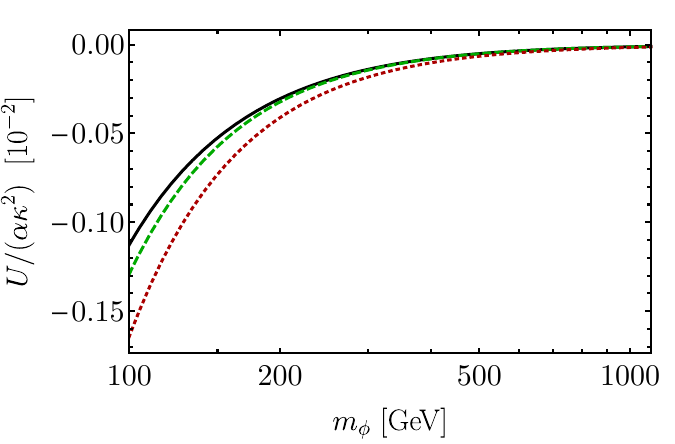} 
\end{center}
\vspace{-4mm} 
\caption{\label{fig:Ucomp} Comparison of the full two-loop result for $U$ with the approximations given in~\eqref{eq:Ulimit}, as a function of the singlet mass $m_\phi$. The solid black curve shows the full result, while the dashed green and dotted red curves correspond to the approximations given in the second and third lines of~\eqref{eq:Ulimit}, respectively. All results are normalised to $\alpha\hspace{0.25mm}\kappa^2$ and given in units of $10^{-2}$.}
\end{figure}

In this appendix, we complete the discussion of Section~\ref{sec:fulltwo} by computing the two-loop corrections to the third Peskin-Takeuchi parameter in the singlet-scalar model. In the conventions of~\eqref{eq:defT} and~\eqref{eq:defS}, it is defined as~\cite{Peskin:1990zt,Peskin:1991sw}
\beq
U = \frac{4 \hspace{0.125mm} s_w^2}{\alpha} \, \Big [ \Pi^\prime_{WW}(0) - c_w^2\,\Pi^\prime_{ZZ}(0) - 2\hspace{0.25mm} s_w \hspace{0.125mm} c_w\,\Pi^\prime_{\gamma Z}(0) - s_w^2\,\Pi^\prime_{\gamma\gamma}(0) \Big ] \,.
\label{eq:defU}
\eeq
Since the photon self-energies receive no $\mathcal{O}(\kappa^2)$ corrections, only the derivatives~\eqref{eq:PiVVprime0} of the $W$- and $Z$-boson self-energies contribute. Expanding them as in~\eqref{eq:Tlimit} and~\eqref{eq:Slimit}, we obtain
\begin{align}
U^{(2)} & = \frac{4 \hspace{0.125mm} s_w^2}{\alpha} \left[ \hat \Pi^{\prime \hspace{0.125mm} (2)}_{WW}(0) - c_w^2 \hspace{0.125mm} \hat \Pi^{\prime \hspace{0.125mm} (2)}_{ZZ}(0) \right] \nonumber \\[1mm]
& \simeq \frac{s_w^2 \hspace{0.25mm} \kappa^2\hspace{0.25mm} m_h^2}{64\hspace{0.125mm}\pi^4\hspace{0.25mm}\alpha\hspace{0.25mm}m_\phi^2} \left[ - \frac{r_W\left(r_W - r_Z\right)\left(8\hspace{0.25mm}r_W r_Z - 3\hspace{0.25mm}r_W - 3\hspace{0.25mm}r_Z - 2\right)}{18\left(r_W - 1\right)^2\left(r_Z - 1\right)^2} \right. \nonumber \\[1mm]
& \hspace{3.5cm} \left. - \frac{r_W^3\left(r_W + 9\right)\ln r_W}{36\left(r_W - 1\right)^3} + \frac{r_W\hspace{0.25mm}r_Z^2\left(r_Z + 9\right)\ln r_Z}{36\left(r_Z - 1\right)^3} \right] \nonumber \\[1mm]
& \simeq -\frac{s_w^2 \hspace{0.25mm} \kappa^2\hspace{0.25mm} v^2}{4608 \hspace{0.125mm}\pi^3\hspace{0.25mm}c_w^2\hspace{0.25mm}m_\phi^2} \,, \label{eq:Ulimit} 
\end{align}
where, as before, the last line corresponds to $m_h = m_Z$ and only the leading term in $s_w^2$ has been kept. Like $T^{(2)}$, the correction~\eqref{eq:Ulimit} vanishes for $r_W = r_Z$, as required for a measure of custodial-symmetry breaking. Unlike $T^{(2)}$ and $S^{(2)}$, however, it contains no logarithm $\ln\hspace{0.125mm}(m_\phi/m_h)$ and is further suppressed by a factor of $s_w^2$. The absence of a logarithmically enhanced term reflects the fact that the SMEFT contains no dimension-six operator that contributes to $U$ at tree level. Consequently, there is no counterterm that could absorb a UV divergence, and the correction~\eqref{eq:Ulimit}, which arises from one-loop matrix elements of~$Q_{H\Box}$, must be finite and free of logarithms. Note that it is nevertheless suppressed by only a single power of $v^2/m_\phi^2$, since it receives contributions from one-loop matrix elements of~$Q_{H\Box}$ evaluated in the broken phase.

Figure~\ref{fig:Ucomp} compares the full two-loop result for $U$ with the approximations~\eqref{eq:Ulimit}. For $m_\phi \gtrsim 300\,{\rm GeV}$, both approximations reproduce the full result well, whereas for lighter singlets they overestimate its magnitude. Numerically, we find $U^{(2)} \simeq -8.7 \cdot 10^{-6} \hspace{0.25mm} \kappa^2$ at $m_\phi = 100\,{\rm GeV}$ and $U^{(2)} \simeq -1.0 \cdot 10^{-7} \hspace{0.25mm} \kappa^2$ at $m_\phi = 1\,{\rm TeV}$. More importantly, a comparison with Figure~\ref{fig:TScomp} shows that $U$ is one to two orders of magnitude smaller than $T$ and $S$ over the entire mass range considered, with the suppression increasing with $m_\phi$. Its impact on EWPO is therefore negligible, which justifies neglecting $U$ in the analysis presented in the main text.

Beyond the Peskin-Takeuchi parameters, universal new physics can modify the gauge-boson self-energies at ${\cal O}(q^4)$. These effects are captured by the parameters $W$ and $Y$ introduced in~\cite{Barbieri:2004qk}. Expressed in terms of the self-energies of the mass eigenstates and in the conventions of~\eqref{eq:defT} and~\eqref{eq:defS}, they read
\beq
\begin{split}
W & = -\frac{m_W^2}{2} \, \Big[ c_w^2 \hspace{0.25mm} \Pi^{\prime\prime}_{ZZ}(0) + 2 \hspace{0.25mm} s_w \hspace{0.125mm} c_w \hspace{0.25mm} \Pi^{\prime\prime}_{\gamma Z}(0) + s_w^2 \hspace{0.25mm} \Pi^{\prime\prime}_{\gamma\gamma}(0) \Big] \,, \\[1mm]
Y & = -\frac{m_W^2}{2} \, \Big[ s_w^2 \hspace{0.25mm} \Pi^{\prime\prime}_{ZZ}(0) - 2 \hspace{0.25mm} s_w \hspace{0.125mm} c_w \hspace{0.25mm} \Pi^{\prime\prime}_{\gamma Z}(0) + c_w^2 \hspace{0.25mm} \Pi^{\prime\prime}_{\gamma\gamma}(0) \Big] \,,
\end{split}
\label{eq:defWY}
\eeq
where the primes denote second derivatives with respect to $q^2$. The combinations in square brackets are the self-energies of the $W^3$ and $B$ fields, in terms of which $W$ and $Y$ were originally defined, and the overall minus sign accounts for the opposite sign convention for the self-energies used in~\cite{Barbieri:2004qk}.

\begin{figure}[t!]
\begin{center}
\includegraphics[width=0.6\textwidth]{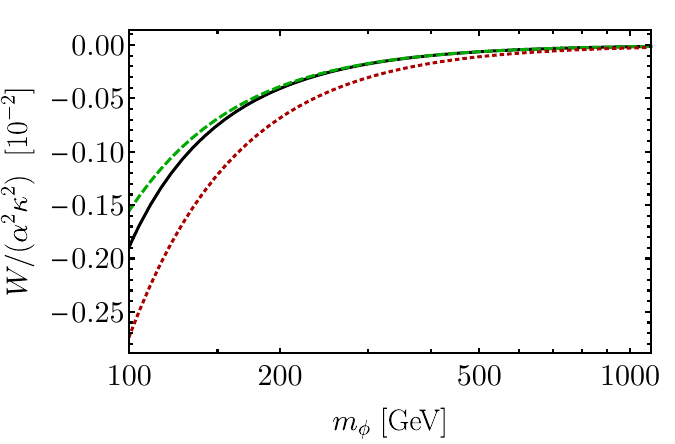}
\end{center}
\vspace{-4mm}
\caption{\label{fig:Wcomp} Comparison of the full two-loop result for $W$ with the approximations given in~\eqref{eq:Wlimit}, as a function of the singlet mass $m_\phi$. The solid black curve shows the full result, while the dashed green and dotted red curves correspond to the approximations given in the second and third lines of~\eqref{eq:Wlimit}, respectively. All results are normalised to $\alpha^2 \kappa^2$ and given in units of $10^{-2}$. The corresponding results for $Y$ follow by multiplication with $s_w^2/c_w^2$.}
\end{figure}

Since the photon self-energies receive no $\mathcal{O}(\kappa^2)$ corrections, only the second derivative of the $Z$-boson self-energy contributes, so that $Y^{(2)} = s_w^2/c_w^2 \, W^{(2)}$ holds exactly at this~order. As in the case of $T$, $S$, and $U$, the singlet enters $\hat \Pi_{ZZ}(q^2)$ only through the renormalised Higgs self-energy, inserted into the internal Higgs line of the one-loop diagrams with a $Z$~boson or a neutral would-be Goldstone boson. In analogy to~\eqref{eq:PiVV0} and~\eqref{eq:PiVVprime0}, we find for~$V = W, Z$
\begin{align}
\hat{\Pi}_{VV}^{\prime\prime \hspace{0.125mm} (2)}(0) &= \frac{\kappa^2}{256\hspace{0.25mm}\pi^4 \hspace{0.25mm} m_\phi^2} \, \Bigg\{ 
- \frac{x_V \hspace{0.25mm} p_1}{24\left(x_h - x_V\right)^5\left(x_V - 4\right)^2} 
+ \frac{x_V^3 \hspace{0.25mm} p_2}{\left(x_h - x_V\right)^6} \ln x_h \nonumber \\[1mm]
& \hspace{5mm} + \frac{x_V \hspace{0.25mm} p_3}{12\left(x_h - x_V\right)^6\left(x_V - 4\right)^3} \ln x_V
- \frac{x_V^3 \hspace{0.25mm} p_4}{2\hspace{0.25mm}x_h\left(x_h - x_V\right)^6}\,\phi\!\left(\frac{x_h}{4}\right) \nonumber \\[1mm]
& \hspace{5mm} + \frac{p_5}{2\left(x_h - x_V\right)^6\left(x_V - 4\right)^3}\,\phi\!\left(\frac{x_V}{4}\right) \nonumber \\[1mm]
& \hspace{5mm} + \left[ \frac{x_V \hspace{0.25mm} p_6}{12\left(x_h - x_V\right)^5} - \frac{x_V^3 \hspace{0.25mm} p_7}{\left(x_h - x_V\right)^6} \, \ln \frac{x_h}{x_V} \right] \Big[ \mathrm{Re} \hspace{0.25mm} B(x_h) - \ln x_\mu \Big] \Bigg\} \,, \label{eq:PiVVpp0}
\end{align}
with
\begin{align}
p_1 & = x_h^3\left(11\hspace{0.25mm}x_V^2 - 42\hspace{0.25mm}x_V + 132\right) - x_h^2 x_V\left(73\hspace{0.25mm}x_V^2 - 334\hspace{0.25mm}x_V + 588\right) \nonumber \\[1mm]
& \phantom{=} \; + x_h x_V^2\left(941\hspace{0.25mm}x_V^2 - 7166\hspace{0.25mm}x_V + 14028\right) + x_V^3\left(801\hspace{0.25mm}x_V^2 - 6566\hspace{0.25mm}x_V + 13308\right) , \nonumber \\[2mm]
p_2 & = 21\hspace{0.25mm}x_h^2 + 45\hspace{0.25mm}x_h x_V + 4\hspace{0.25mm}x_V^2 \,, \nonumber \\[2mm]
p_3 & = x_h^4\left(x_V^3 + 6\hspace{0.25mm}x_V^2 + 98\hspace{0.25mm}x_V - 132\right) + 4\hspace{0.25mm}x_h^3 x_V\left(x_V^3 - 40\hspace{0.25mm}x_V^2 - 6\hspace{0.25mm}x_V + 180\right) \nonumber \\[1mm]
& \phantom{=} \; + 12\hspace{0.25mm}x_h^2 x_V^2\left(40\hspace{0.25mm}x_V^2 - 139\hspace{0.25mm}x_V + 126\right) - 8\hspace{0.25mm}x_h x_V^3\left(80\hspace{0.25mm}x_V^3 - 873\hspace{0.25mm}x_V^2 + 3367\hspace{0.25mm}x_V - 4410\right) \nonumber \\[1mm]
& \phantom{=} \; - 5\hspace{0.25mm}x_V^4\left(41\hspace{0.25mm}x_V^3 - 554\hspace{0.25mm}x_V^2 + 2358\hspace{0.25mm}x_V - 3276\right) , \nonumber \\[2mm]
p_4 & = 9\hspace{0.25mm}x_h^3 + 2\hspace{0.25mm}x_h^2\left(11\hspace{0.25mm}x_V - 21\right) + 2\hspace{0.25mm}x_h x_V\left(2\hspace{0.25mm}x_V - 45\right) - 8\hspace{0.25mm}x_V^2 \,, \nonumber \\[2mm]
p_5 & = -2\hspace{0.25mm}x_h^4\left(4\hspace{0.25mm}x_V^2 - 10\hspace{0.25mm}x_V + 11\right) + 8\hspace{0.25mm}x_h^3 x_V\left(6\hspace{0.25mm}x_V^2 - 19\hspace{0.25mm}x_V + 15\right) \nonumber \\[1mm]
& \phantom{=} \; + 3\hspace{0.25mm}x_h^2 x_V^2\left(3\hspace{0.25mm}x_V^4 - 42\hspace{0.25mm}x_V^3 + 170\hspace{0.25mm}x_V^2 - 256\hspace{0.25mm}x_V + 84\right) \nonumber \\[1mm]
& \phantom{=} \; + 2\hspace{0.25mm}x_h x_V^3\left(11\hspace{0.25mm}x_V^4 - 166\hspace{0.25mm}x_V^3 + 1018\hspace{0.25mm}x_V^2 - 2820\hspace{0.25mm}x_V + 2940\right) \nonumber \\[1mm]
& \phantom{=} \; + 2\hspace{0.25mm}x_V^4\left(2\hspace{0.25mm}x_V^4 - 51\hspace{0.25mm}x_V^3 + 387\hspace{0.25mm}x_V^2 - 1210\hspace{0.25mm}x_V + 1365\right) , \nonumber \\[2mm]
p_6 & = x_h^3 + 5\hspace{0.25mm}x_h^2 x_V + 257\hspace{0.25mm}x_h x_V^2 + 157\hspace{0.25mm}x_V^3 \,, \nonumber \\[2mm]
p_7 & = 9\hspace{0.25mm}x_h^2 + 22\hspace{0.25mm}x_h x_V + 4\hspace{0.25mm}x_V^2 \,.
\label{eq:ppoly}
\end{align}
In contrast to~\eqref{eq:PiVV0}, the result~\eqref{eq:PiVVpp0} is free of $1/\epsilon$ poles, and, like~\eqref{eq:PiVVprime0}, it depends on the renormalisation scale only through the combination $\mathrm{Re}\hspace{0.25mm}B(x_h) - \ln x_\mu$, which is $\mu$ independent. Expanding~\eqref{eq:PiVVpp0} for $V = Z$ as in~\eqref{eq:Tlimit} and~\eqref{eq:Slimit}, we obtain
\begin{align}
W^{(2)} & = -\frac{c_w^2 \hspace{0.25mm} m_W^2}{2} \, \hat \Pi^{\prime\prime \hspace{0.125mm} (2)}_{ZZ}(0) \nonumber \\[1mm]
& \simeq -\frac{c_w^2 \hspace{0.25mm} \kappa^2 \hspace{0.25mm} m_W^2}{6144 \hspace{0.125mm} \pi^4 \hspace{0.25mm} m_\phi^2} \left[ \frac{r_Z \left( 9 \hspace{0.25mm} r_Z^4 + 64 \hspace{0.25mm} r_Z^3 - 72 \hspace{0.25mm} r_Z^2 - 1 \right)}{6 \left( r_Z - 1 \right)^5} - \frac{2 \hspace{0.25mm} r_Z^3 \left( 4 \hspace{0.25mm} r_Z + 3 \right) \ln r_Z}{\left( r_Z - 1 \right)^5} \right] \nonumber \\[1mm]
& \simeq -\frac{c_w^2 \hspace{0.25mm} \alpha \hspace{0.25mm} \kappa^2 \hspace{0.25mm} v^2}{30720 \hspace{0.125mm} \pi^3 \hspace{0.25mm} s_w^2 \hspace{0.25mm} m_\phi^2} \,, \label{eq:Wlimit}
\end{align}
where the last line corresponds to $m_h = m_Z$. Like $U^{(2)}$, the corrections $W^{(2)}$ and $Y^{(2)}$ contain no logarithm $\ln\hspace{0.125mm}(m_\phi/m_h)$, but for a different reason. The SMEFT does contain dimension-six operators that contribute to $W$ and $Y$ at tree level, namely $O_{2W}$ and $O_{2B}$ given in~\eqref{eq:SILHgauge}. However, $Q_{H\Box}$ does not mix into these operators at one loop, since a gauge-boson two-point function with a single insertion of $Q_{H\Box}$ requires two loops in the unbroken phase. Moreover, their matching corrections first arise at three loops, because the singlet couples to the gauge bosons only through closed Higgs loops. The one-loop matrix elements of $Q_{H\Box}$ that give rise to~\eqref{eq:Wlimit} must therefore be finite and free of logarithms. Indeed, the second line of~\eqref{eq:Wlimit} is proportional to $C_{H\Box}^{(1)}$ given in~\eqref{eq:high}, i.e.~it is entirely due to the rescaling of the Higgs couplings to the $Z$ boson. As for $U^{(2)}$, the corrections are nevertheless suppressed by only a single power of $v^2/m_\phi^2$.

Figure~\ref{fig:Wcomp} compares the full two-loop result for $W$ with the approximations~\eqref{eq:Wlimit}. The~second line of~\eqref{eq:Wlimit} reproduces the full result to better than $5\%$ for $m_\phi \gtrsim 200\,{\rm GeV}$ and underestimates its magnitude by about $20\%$ at $m_\phi = 100\,{\rm GeV}$. The third line overestimates the magnitude by $45\%$ to $75\%$, reflecting the sizeable dependence of the square bracket in~\eqref{eq:Wlimit} on the Higgs-boson mass. Numerically, we find $W^{(2)} \simeq -1.2 \cdot 10^{-7} \hspace{0.25mm} \kappa^2$ at $m_\phi = 100\,{\rm GeV}$ and $W^{(2)} \simeq -1.0 \cdot 10^{-9} \hspace{0.25mm} \kappa^2$ at $m_\phi = 1\,{\rm TeV}$, while $Y^{(2)}$ is smaller by a factor of~$s_w^2/c_w^2 \simeq 0.3$. These corrections lie one to two orders of magnitude below the corresponding ones to $\hat T = \alpha \hspace{0.25mm} T$ and $\hat S = \alpha \hspace{0.25mm} S/(4 \hspace{0.125mm} s_w^2)$, and are similar in size to those to $\hat U = \alpha \hspace{0.25mm} U$. The~parameters $W$ and $Y$ can therefore be safely neglected in the EWPO analysis presented in the main~text.

\section{Details of the two-loop SMEFT matching calculation}
\label{sec:SMEFTdetails}

In this appendix, we describe how the Wilson coefficients $c_T$ and $c_W + c_B$, which determine the Peskin-Takeuchi parameters $T$ and $S$ at tree level, are extracted from off-shell Green's functions in the unbroken phase of the SMEFT. The same method also yields the Wilson coefficient $c_H$ and the coefficient $c_r$ of a redundant operator, which together determine $C_{H\Box}$.

\subsection*{General strategy}

In DR, the matching corrections are given by the hard region of the full-theory diagrams, in which all loop momenta are of the order of the singlet mass~$m_\phi$. The masses of the Higgs and gauge bosons can therefore be set to zero, and the integrands can be Taylor expanded in the external momenta prior to integration. Since the relevant operator structures are of second order in the external momenta, the expansion is carried out to this order. We~consider one-light-particle-irreducible~(1LPI) Green's functions, i.e.~diagrams that cannot be disconnected by cutting a single Higgs or gauge-boson line. The IR poles of the hard-region contributions cancel against the UV poles of the corresponding EFT matrix elements and are removed in the $\overline{\rm MS}$ scheme.

\subsection*{Projection onto $\bm{c_T}$ and $\bm{c_H}$}

The Wilson coefficient $c_T$ is extracted from the four-point function
\beq
H_1(p_1) \hspace{0.5mm} H_2(p_2) \to H_1(-p_3) \hspace{0.5mm} H_2(-p_4) \,,
\label{eq:HHHH}
\eeq
where all momenta are taken to be incoming, i.e.~$\sum_{i=1}^4 p_i = 0$, and the subscripts label the $SU(2)_L$ components of the Higgs doublet. The 1LPI Green's function has the general isospin structure
\beq
\Gamma_{ij}^{kl} = \delta_i^k\hspace{0.25mm}\delta_j^l\hspace{0.5mm} A(p_1, p_2; p_3, p_4) + \delta_i^l\hspace{0.25mm}\delta_j^k\hspace{0.5mm} A(p_1, p_2; p_4, p_3) \,,
\label{eq:Gamma4}
\eeq
where $i$, $j$ and $k$, $l$ are the $SU(2)_L$ indices of the incoming and outgoing Higgs fields, respectively, and the choice of external components in~\eqref{eq:HHHH} selects the first term, so that the process~\eqref{eq:HHHH} directly yields $A = A(p_1, p_2; p_3, p_4)$. Since in the bosonic sector custodial symmetry is broken only by hypercharge interactions, a non-zero contribution to $c_T$ can only arise from diagrams with~$U(1)_Y$ gauge-boson exchange, as we verify explicitly below.

At $\mathcal{O}(p^2)$, the amplitude $A$ receives tree-level contributions from four operators, namely the SILH operators $O_H$ and $O_T$ introduced in~\eqref{eq:SILH} and the two redundant operators
\beq
O_r = |H|^2 \big(D_\mu H\big)^\dagger \big(D^\mu H\big) \,, \qquad
O_r^{\prime\prime} = |H|^2 \hspace{0.25mm} \partial_\mu \big(H^\dagger i \overleftrightarrow{D}^\mu H\big) \,.
\label{eq:redundant}
\eeq
The operator $O_r^{\prime\prime}$ vanishes by the Higgs equation of motion~(EOM). Integrating by parts, one instead finds $O_r = -O_H - 1/2 \hspace{0.5mm} |H|^2 \big[ H^\dagger D^2 H + {\rm h.c.} \big]$, so that eliminating $O_r$ with the EOM shifts the coefficient of $O_H$ to $c_H - c_r$. Neither operator generates $O_T$ in this process. In terms of the invariants
\beq
X_1 = p_1 \cdot p_3 + p_2 \cdot p_4 \,, \quad
X_2 = p_1 \cdot p_4 + p_2 \cdot p_3 \,, \quad
X_3 = p_1^2 + p_2^2 \,, \quad
X_4 = p_3^2 + p_4^2 \,,
\label{eq:Xi}
\eeq
where $a \cdot b = a^\mu b_\mu$ denotes the Minkowski scalar product, the contributions to $A$ from a single tree-level insertion of the operators $c_i\hspace{0.25mm} O_i$ with $i = H, T, r, r^{\prime\prime}$, which we denote by~$A\hspace{0.25mm}\big|_{O_i}$, read
\beq
\begin{gathered}
A\hspace{0.25mm}\big|_{O_H} = c_H \left[ X_1 + \frac{1}{2} \left( X_3 + X_4 \right) \right] , \qquad 
A\hspace{0.25mm}\big|_{O_T} = c_T \left[ X_1 + 2\hspace{0.25mm} X_2 + \frac{1}{2} \left( X_3 + X_4 \right) \right] , \\[2mm]
A\hspace{0.25mm}\big|_{O_r}  = -c_r \hspace{0.5mm} X_1 \,, \qquad 
A\hspace{0.25mm}\big|_{O_r^{\prime\prime}} = -i\hspace{0.25mm} c_r^{\prime\prime} \left( X_3 - X_4 \right) .
\end{gathered}
\label{eq:Atree}
\eeq 
Here, the amplitude is normalised such that it coincides with the coefficient of the corresponding field monomial in the Lagrangian after the replacement $\partial_\mu \to -i\hspace{0.25mm}p_\mu$ for all fields, with all momenta incoming. Only $O_T$ generates the structure $X_2$, which corresponds to contracting the momenta of $H$ and $H^\dagger$ fields that carry different $SU(2)_L$ indices and therefore violates custodial symmetry. After eliminating $p_4$ by momentum conservation, the amplitude at $\mathcal{O}(p^2)$ is a quadratic form in the independent momenta,
\beq
A = \sum_{a \leq b} \hspace{0.5mm} m_{ab} \hspace{0.75mm} p_{a} \cdot p_{b} \,,
\label{eq:quadform}
\eeq
where the indices $a$ and $b$ run over $1, 2, 3$. From~\eqref{eq:Atree}, it follows that the combinations
\beq
c_T = \frac{1}{4} \left( m_{23} - m_{12} \right) , \qquad
c_H = m_{11} + \frac{1}{4} \left( m_{23} - m_{12} \right) ,
\label{eq:projectors}
\eeq
project onto the Wilson coefficients $c_T$ and $c_H$, respectively, while the contributions of the redundant operators drop out. The coefficient of $O_r$ is given by $c_r = m_{22}$, so that the physical combination $c_H - c_r$ is also determined. Equivalently, the projectors can be written in terms of derivatives with respect to the external momenta,
\beq
\begin{gathered}
c_T = \frac{1}{4\hspace{0.25mm}d} \left( \frac{\partial}{\partial p_2^\mu} \frac{\partial}{\partial p_{3\mu}} - \frac{\partial}{\partial p_1^\mu} \frac{\partial}{\partial p_{2\mu}} \right) A \hspace{0.5mm} \bigg|_{p_i = 0} \,, \qquad
c_H = \frac{1}{2\hspace{0.25mm}d} \hspace{0.5mm} \frac{\partial}{\partial p_1^\mu} \frac{\partial}{\partial p_{1\mu}} A \hspace{0.5mm} \bigg|_{p_i = 0} + c_T \,, \\[2mm]
c_r = \frac{1}{2\hspace{0.25mm}d} \hspace{0.5mm} \frac{\partial}{\partial p_2^\mu} \frac{\partial}{\partial p_{2\mu}} A \hspace{0.5mm} \bigg|_{p_i = 0} \,,
\end{gathered}
\label{eq:projectorsD}
\eeq
where $A$ is expressed in terms of $p_1$, $p_2$, and $p_3$ before differentiating, and $d = 4 - 2\hspace{0.125mm}\epsilon$. As~a check of the normalisation, we have verified that applying~\eqref{eq:projectorsD} to the one-loop singlet contribution yields $c_r^{(1)} = c_T^{(1)} = 0$ and reproduces $c_H^{(1)} = -2\hspace{0.25mm}C_{H\Box}^{(1)}$, with $C_{H\Box}^{(1)}$ given in~\eqref{eq:high}. We have furthermore verified explicitly that the two-loop diagrams with $W$-boson exchange do not contribute to $c_T^{(2)}$, as required by custodial symmetry.

\subsection*{Projection onto $\bm{c_W + c_B}$}

Besides $O_W$ and $O_B$, the operators relevant for the $W$--$B$ sector are
\beq
\begin{gathered}
O_{HW} = i\hspace{0.25mm} g_L \big(D^\mu H\big)^\dagger \sigma^a \big(D^\nu H\big) W^a_{\mu\nu} \,, \qquad
O_{HB} = i\hspace{0.25mm} g_Y \big(D^\mu H\big)^\dagger \big(D^\nu H\big) B_{\mu\nu} \,, \\[2mm]
O_{WW} = g_L^2 \hspace{0.25mm} |H|^2 W^a_{\mu\nu} W^{a\mu\nu} \,, \qquad
O_{BB} = g_Y^2 \hspace{0.25mm} |H|^2 B_{\mu\nu} B^{\mu\nu} \,, \\[2mm]
O_{WB} = g_L\hspace{0.25mm} g_Y \big(H^\dagger \sigma^a H\big) W^a_{\mu\nu} B^{\mu\nu} \,, \\[2mm]
O_{2W} = -\frac{1}{2} \left( D^\mu W^a_{\mu\nu} \right)^2 , \qquad
O_{2B} = -\frac{1}{2} \left( \partial^\mu B_{\mu\nu} \right)^2 .
\end{gathered}
\label{eq:SILHgauge}
\eeq
Up to total derivatives, these operators satisfy the identities~\cite{Giudice:2007fh,Elias-Miro:2013mua}
\beq
O_W = O_{HW} + \frac{1}{4} \hspace{0.25mm} O_{WW} + \frac{1}{4} \hspace{0.25mm} O_{WB} \,, \qquad
O_B = O_{HB} + \frac{1}{4} \hspace{0.25mm} O_{BB} + \frac{1}{4} \hspace{0.25mm} O_{WB} \,,
\label{eq:SILHidentities}
\eeq
which are used in the SILH basis to eliminate $O_{WW}$ and $O_{WB}$. Since the two sides of these identities differ only by operators with two field-strength tensors, $O_W$ and $O_{HW}$ induce identical $H H^\dagger W$ vertices at $\mathcal{O}(p^3)$, and so do $O_B$ and $O_{HB}$ for $H H^\dagger B$. The three-point functions $H H^\dagger W$ and $H H^\dagger B$ therefore cannot disentangle $O_W$ and $O_B$ from $O_{HW}$ and~$O_{HB}$.

The $S$ parameter, however, only depends on the mixing between the $W^3$ and $B$ bosons in a constant Higgs background. In such a background, the covariant derivative reduces to $D_\mu H = -i \hspace{0.25mm} \mathcal{A}_\mu H$ with $\mathcal{A}_\mu = g_L/2 \hspace{0.5mm} \sigma^a W^a_\mu + g_Y/2 \hspace{0.5mm} B_\mu$. Since $\big(D^\mu H\big)^\dagger \big(D^\nu H\big) = H^\dagger \mathcal{A}^\mu \mathcal{A}^\nu H$, the operators $O_{HW}$ and $O_{HB}$ then contain at least three gauge fields and do not contribute to $W^3$--$B$ mixing. The operators $O_{WW}$ and $O_{BB}$ only renormalise the gauge-boson kinetic terms. The operators $O_{2W}$ and $O_{2B}$ contain no Higgs fields, do not mix $W^3$ and $B$, and enter the gauge-boson self-energies only at $\mathcal{O}(k^4)$, i.e.~through the $W$ and $Y$ parameters~\cite{Barbieri:2004qk}. Operators without derivatives acting on gauge fields contribute at most at~$\mathcal{O}(k^0)$. In~contrast, using $H^\dagger \sigma^a i \overleftrightarrow{D}_\mu H = H^\dagger \{ \sigma^a, \mathcal{A}_\mu \} H$, where the curly brackets denote the anticommutator, and $H^\dagger i \overleftrightarrow{D}_\mu H = 2 \hspace{0.25mm} H^\dagger \mathcal{A}_\mu H$ and integrating by parts, one finds that $c_W\hspace{0.25mm} O_W + c_B\hspace{0.25mm} O_B$ contains the term
\beq
X \hspace{0.25mm} \big(H^\dagger \sigma^a H\big) \hspace{0.25mm} W^a_{\mu\nu} B^{\mu\nu} \,, \qquad
X = \frac{g_L\hspace{0.25mm} g_Y}{4} \left( c_W + c_B \right) ,
\label{eq:Xdef}
\eeq
in agreement with~\eqref{eq:SILHidentities}. The only other contribution arises from the redundant operator $\big(D^2 H\big)^\dagger \big(D^2 H\big)$. Since $D^2 H \supset -i \hspace{0.25mm} \big(\partial \cdot \mathcal{A}\big) H$, it only generates terms proportional to $k^\mu k^\nu$ in momentum space.

We therefore consider the four-point function
\beq
H_1(0) \hspace{0.5mm} H_1^\dagger(0) \to W^3(k) \hspace{0.5mm} B(-k) \,,
\label{eq:HHWB}
\eeq
with vanishing Higgs momenta. Since $\big(\sigma^a\big)_{11} = \delta^{a3}$, the operator~\eqref{eq:Xdef} couples $H_1 H_1^\dagger$ only to $W^3$, and at tree level it gives
\beq
A = 2\hspace{0.25mm} X \left[ k^2 \left( \varepsilon_W \cdot \varepsilon_B \right) - \left( k \cdot \varepsilon_W \right) \left( k \cdot \varepsilon_B \right) \right] ,
\label{eq:AWB}
\eeq
where $\varepsilon_W$ and $\varepsilon_B$ denote the polarisation vectors of the gauge bosons. Since the redundant operator $\big(D^2 H\big)^\dagger \big(D^2 H\big)$ also contributes to the coefficient of $\left( k \cdot \varepsilon_W \right) \left( k \cdot \varepsilon_B \right)$, the hard-region result need not have the transverse form~\eqref{eq:AWB}. The coefficient $X$, and hence the combination $c_W + c_B$, is therefore obtained as one half of the coefficient of $k^2 \left( \varepsilon_W \cdot \varepsilon_B \right)$ in the $\mathcal{O}(k^2)$ part of the four-point function~\eqref{eq:HHWB}, i.e.~$c_W + c_B = 4 X/(g_L \hspace{0.25mm} g_Y)$.

\subsection*{Momentum expansion and reduction}

We denote the two loop momenta by $\ell_1$ and $\ell_2$. The momentum flowing through each propagator can be written as $\ell + P$, where $\ell$ is a linear combination of $\ell_1$ and $\ell_2$ and $P$ is a linear combination of the external momenta. Each propagator is then expanded to second order in $P$ as
\beq
\frac{1}{(\ell + P)^2 - m^2} = \frac{1}{\ell^2 - m^2} - \frac{2\hspace{0.25mm}\ell \cdot P + P^2}{\left( \ell^2 - m^2 \right)^2} + \frac{4 \left( \ell \cdot P \right)^2}{\left( \ell^2 - m^2 \right)^3} + \mathcal{O}(P^3) \,.
\label{eq:propexp}
\eeq
After the expansion, all loop integrals are vacuum integrals, and the external momenta and polarisation vectors appear only in scalar products with $\ell_1$ and $\ell_2$ or with each other. Tensor integrals of odd rank vanish, while those of rank two and four are reduced using
\beq
\begin{gathered}
\left( \ell_{a_1} \cdot x_1 \right) \left( \ell_{a_2} \cdot x_2 \right) \to \frac{1}{d} \left( \ell_{a_1} \cdot \ell_{a_2} \right) \left( x_1 \cdot x_2 \right) , \\[2mm]
\prod_{i=1}^4 \left( \ell_{a_i} \cdot x_i \right) \to \sum_{\pi} \frac{1}{d \left( d - 1 \right)} \left[ s_\pi - \frac{1}{d+2} \sum_{\pi^\prime} s_{\pi^\prime} \right] e_\pi \,,
\end{gathered}
\label{eq:tensorred}
\eeq
where $a_i \in \{1, 2\}$, the vectors $x_i$ denote external momenta or polarisation vectors, and $\pi$ and $\pi^\prime$ run over the three pairings $(12)(34)$, $(13)(24)$, and $(14)(23)$ of the indices $i = 1, 2, 3, 4$. For $\pi = (12)(34)$, for instance, $s_\pi = (\ell_{a_1} \cdot \ell_{a_2})(\ell_{a_3} \cdot \ell_{a_4})$ and $e_\pi = (x_1 \cdot x_2)(x_3 \cdot x_4)$. Rank-four tensors only occur in the four-point function~\eqref{eq:HHWB}, where both polarisation vectors and two powers of $k$ can be contracted with loop momenta. In the four-point function~\eqref{eq:HHHH}, the expansion to second order in the external momenta generates at most rank-two tensors.

The resulting scalar two-loop vacuum integrals involve propagators with mass $m_\phi$ and massless propagators. After a suitable linear transformation of the loop momenta, the propagators of each diagram can be grouped into three chains carrying the momenta $\ell_1$, $\ell_2$, and $\ell_1 - \ell_2$. Scalar products in the numerator are expressed in terms of these propagators, and propagators in the same chain but with different masses are separated by partial fractioning. The resulting integrals are reduced to master integrals using integration-by-parts~\cite{Tkachov:1981wb,Chetyrkin:1981qh} and dimensional recurrence relations~\cite{Tarasov:1996br,Tarasov:1997kx}, as implemented in the in-house \texttt{Mathematica} package \texttt{Coltrane}~\cite{Haisch:2002zz}. The master integrals are one- and two-loop vacuum integrals with one or two propagators of mass $m_\phi$, which are known in closed form in $d = 4 - 2\hspace{0.125mm}\epsilon$ dimensions~\cite{Davydychev:1992mt}.

\section{Dispersive representation of the gauge-boson self-energies}
\label{app:dispersive}

In the following, we derive a compact representation of the two-loop corrections to the gauge-boson self-energies in the singlet-scalar model, which holds for arbitrary $q^2$ and to all orders in $v^2/m_\phi^2$. It provides an independent check of the results presented in Section~\ref{sec:fulltwo} and Appendix~\ref{app:UWY}, makes the origin of the IR logarithms discussed in Section~\ref{sec:IRlogs} transparent, and applies to a broad class of hidden sectors that couple to the SM through the operator~$|H|^2$.

As discussed in Section~\ref{sec:fulltwo}, at~${\cal O}(\kappa^2)$ the singlet enters the renormalised self-energies $\hat \Pi_{VV}(q^2)$ with $V = W, Z$ only through the singlet-loop correction to the Higgs-boson propagator. The terms proportional to ${\rm Re}\hspace{0.25mm}A_0(m_\phi^2)$ cancel against the corresponding counterterms, and the Higgs-boson tadpoles are removed by the tadpole counterterm. We write the dressed Higgs-boson propagator as $i/\big(k^2 - m_h^2 + \hat \Sigma_{hh}(k^2)\big)$, where $k$ denotes the momentum flowing through the propagator, and use the conventions of~\cite{Hahn:1998yk,Hahn:2016ebn} that also underlie~\eqref{eq:dZh} and~\eqref{eq:dmh2}. The trilinear vertex $-\kappa\hspace{0.25mm}v\hspace{0.25mm}h\hspace{0.25mm}\phi^2/2$, which follows from~\eqref{eq:Lphi} after EW symmetry breaking, then induces the renormalised self-energy
\beq
\hat \Sigma_{hh}(k^2) = \frac{\kappa^2 v^2}{32\hspace{0.125mm}\pi^2} \, \Big[ B_0(k^2, m_\phi^2, m_\phi^2) - {\rm Re}\hspace{0.25mm} B_0(m_h^2, m_\phi^2, m_\phi^2) \Big] \,,
\label{eq:Sigmahh}
\eeq
where the subtraction at $k^2 = m_h^2$ implements the OS renormalisation of the Higgs-boson mass, cf.~\eqref{eq:dmh2}. For $m_\phi > m_h/2$, the function~\eqref{eq:Sigmahh} is analytic in the complex $k^2$ plane except for a branch cut along the real axis at $k^2 = \hat s \geq 4\hspace{0.25mm}m_\phi^2$, across which ${\rm Im}\hspace{0.25mm}B_0(\hat s, m_\phi^2, m_\phi^2) = \pi \hspace{0.25mm} \beta(\hat s)$ with $\beta(\hat s) = \sqrt{1 - 4\hspace{0.25mm}m_\phi^2/\hat s}$. Representing $\hat \Sigma_{hh}(k^2)$ by a once-subtracted dispersion integral and expanding the dressed propagator to first order in $\hat \Sigma_{hh}(k^2)$, we obtain
\beq
-\frac{\hat \Sigma_{hh}(k^2)}{\left(k^2 - m_h^2\right)^2} = \int_{4 m_\phi^2}^\infty \! d\hat s \, \rho(\hat s) \left[  \frac{1}{k^2 - \hat s} - \frac{1}{k^2 - m_h^2} \right] , \qquad
\rho(\hat s) = \frac{\kappa^2 v^2}{32\hspace{0.125mm}\pi^2} \, \frac{\beta(\hat s)}{\left(\hat s - m_h^2\right)^2} \,.
\label{eq:propdisp}
\eeq
At~${\cal O}(\kappa^2)$, the singlet thus adds to the Higgs-boson propagator a continuum of scalar states with masses $\sqrt{\hat s} \geq 2\hspace{0.25mm}m_\phi$ and positive spectral weight~$\rho(\hat s)$, while it reduces the residue of the Higgs-boson pole by
\beq
\int_{4 m_\phi^2}^\infty \! d\hat s \, \rho(\hat s) = 1 - \kappa_V^2 = \frac{\kappa^2 v^2}{192\hspace{0.125mm}\pi^2 \hspace{0.25mm} m_\phi^2} \left[ 1 + {\cal O}\left(\frac{m_h^2}{m_\phi^2}\right) \right] .
\label{eq:sumrule}
\eeq
The leading term in~\eqref{eq:sumrule} agrees with $1 - \kappa_V^2 \simeq -2\hspace{0.25mm}v^2\hspace{0.125mm}C_{H\Box}^{(1)}$ and~\eqref{eq:high}.

Since the Higgs-boson propagator enters the two-loop diagrams of Figure~\ref{fig:diagramsk2} linearly, the decomposition~\eqref{eq:propdisp} carries over to the gauge-boson self-energies. One obtains
\beq
\hat \Pi^{(2)}_{VV}(q^2) = \int_{4 m_\phi^2}^\infty \! d \hat s \, \rho(\hat s) \, \Big[ \Pi^h_{VV}(q^2; \hat s) - \Pi^h_{VV}(q^2; m_h^2) \Big] \,,
\label{eq:Pidisp}
\eeq
where $\Pi^h_{VV}(q^2; m^2)$ denotes the sum of the one-loop contributions to the SM self-energy $\Pi_{VV}(q^2)$ that contain an internal Higgs-boson line, evaluated for a Higgs-boson mass~$m$. In the conventions of~\eqref{eq:defT} and~\eqref{eq:defS} and in the Feynman gauge of the background-field formalism, in which the relevant vertices coincide with those of the conventional Feynman gauge, the diagrams with a Higgs boson and a gauge boson, with a Higgs boson and a would-be Goldstone boson, and with a Higgs-boson loop attached to a quartic seagull vertex give
\beq
\Pi^h_{VV}(q^2; m^2) = -\frac{m_V^2}{16\hspace{0.125mm}\pi^2 \hspace{0.25mm} v^2} \, \Big[ 4\hspace{0.25mm} m_V^2 \hspace{0.25mm} B_0(q^2, m^2, m_V^2) - 4 \hspace{0.25mm} B_{00}(q^2, m^2, m_V^2) + A_0(m^2) \Big] \,,
\label{eq:Pih}
\eeq
with the standard one-loop integrals $A_0(m^2)$, $B_0(q^2, m^2, m_V^2)$, and $B_{00}(q^2, m^2, m_V^2)$ in the conventions of~\cite{Hahn:1998yk,Hahn:2016ebn}. The UV poles of $A_0(m^2)$ and $B_{00}(q^2, m^2, m_V^2)$ that depend on $m$ cancel in~\eqref{eq:Pih}, and the remaining poles and the dependence on $\mu$ drop out of the difference in~\eqref{eq:Pidisp}. For large $\hat s$, however, $\Pi^h_{VV}(q^2; \hat s) \simeq m_V^2 \hspace{0.25mm} \hat s/(32 \pi^2 v^2)$ independently of $q^2$, so that $\hat \Pi_{VV}(q^2)$ contains the term $m_V^2/(32 \pi^2 v^2) \int \! d\hat s \, \rho(\hat s) \, \hat s$, which diverges logarithmically even though~\eqref{eq:sumrule} converges. Since its coefficient is the same for $V = W, Z$, this term amounts to a universal shift of~$v^2$. It corresponds to the residual pole in~\eqref{eq:PiVV0} and cancels in~$T$. The derivatives $\hat\Pi^{\prime}_{VV}(0)$ and $\hat\Pi^{\prime\prime}_{VV}(0)$, and hence $S$, $U$, $W$, and $Y$, are finite and scale independent. Expanding~\eqref{eq:Pidisp} in $q^2$ and performing the integral over~$\hat s$ numerically, we reproduce~\eqref{eq:PiVVprime0} and~\eqref{eq:PiVVpp0} exactly, as well as~\eqref{eq:PiVV0} in the combination that enters~$T$. This constitutes an independent check of our full two-loop calculation in the broken~EW~phase.

The representation~\eqref{eq:Pidisp} separates the two sources of the corrections in a transparent~way. The second term in the square bracket is the one-loop SM contribution multiplied by $-(1 - \kappa_V^2)$, i.e.~it describes the rescaling of the Higgs couplings to the EW gauge bosons and, at leading order in $v^2/m_\phi^2$, corresponds to the one-loop matrix elements of $Q_{H\Box}$ discussed in~Section~\ref{sec:SMEFT}. The first term arises from the continuum at $\hat s \geq 4\hspace{0.25mm}m_\phi^2$ and contains the effects of hard loop momenta of the order of~$m_\phi$. The large-$\hat s$ behaviour of the functions~\eqref{eq:Pih} then determines the structure of the oblique parameters. For $\hat s \gg m_V^2$, one finds for instance
\beq
\Pi^{h \hspace{0.125mm} \prime}_{VV}(0; \hat s) = \frac{m_V^2}{48\hspace{0.125mm}\pi^2 \hspace{0.25mm} v^2} \, \ln \frac{\hat s}{\mu^2} + {\cal O}(\hat s^0) \,, \qquad
\Pi^{h \hspace{0.125mm} \prime\prime}_{VV}(0; \hat s) = {\cal O}\left( \frac{1}{\hat s} \right) .
\label{eq:largeS}
\eeq
With~\eqref{eq:sumrule}, the logarithm in the first relation turns into $(1 - \kappa_V^2) \ln \hspace{0.125mm} (m_\phi^2/m_h^2)$ up to non-logarithmic terms, which reproduces the logarithmically enhanced term of $S$ in~\eqref{eq:IRlogs}. The~same mechanism generates the logarithm in~$T$. In contrast, the continuum contribution to $\hat \Pi^{\prime\prime}_{VV}(0)$ is suppressed by an additional power of $v^2/m_\phi^2$, so that the leading terms of $W$ and $Y$ are entirely due to the rescaling of the Higgs couplings and contain no logarithm, in agreement with the discussion in Appendix~\ref{app:UWY}. Since~\eqref{eq:Pidisp} holds for all $q^2$, it also provides the full momentum dependence of the corrections, including the region $q^2 \gtrsim 4\hspace{0.25mm}m_\phi^2$ in which an expansion in oblique parameters is no longer adequate. Analogous form factors have been used to probe EW multiplets in high-energy Drell-Yan production~\cite{DiLuzio:2018jwd}, where the full momentum dependence turns out to be important close to the pair-production threshold. In the singlet-scalar model, the resulting effects are, however, suppressed by~$1 - \kappa_V^2$ and an EW loop factor, and they are too small to be observable.

Finally, the derivation of~\eqref{eq:Pidisp} does not rely on the specific form of the singlet sector. Consider a hidden sector that couples to the SM only through ${\cal L} \supset -|H|^2 \hspace{0.25mm} {\cal O}$, where ${\cal O}$ is a gauge-singlet operator, and assume that the Higgs boson does not mix with the hidden-sector states. At second order in the portal interaction,~\eqref{eq:Pidisp} then continues to hold with
\beq
\begin{gathered}
\rho(\hat s) = \frac{v^2 \hspace{0.25mm} \rho_{\cal O}(\hat s)}{\left(\hat s - m_h^2\right)^2} \,, \qquad
\rho_{\cal O}(\hat s) = \frac{1}{\pi} \hspace{0.5mm} {\rm Im} \hspace{0.25mm} \Pi_{\cal O}(\hat s) \,, \\[2mm]
\Pi_{\cal O}(p^2) = i \int \! d^4x \, e^{i p \cdot x} \hspace{0.25mm} \langle 0 | T\big[ {\cal O}(x) \hspace{0.25mm} {\cal O}(0) \big] | 0 \rangle \,,
\end{gathered}
\label{eq:rhogeneral}
\eeq
where $T$ denotes time ordering and the integration in~\eqref{eq:Pidisp} now starts at the lowest threshold of the hidden sector. For ${\cal O} = \kappa \hspace{0.25mm} \phi^2/2$, one recovers~\eqref{eq:propdisp}. The oblique parameters $T$, $S$, $U$, $W$, and $Y$ of any such sector are thus determined by its spectral function~$\rho_{\cal O}(\hat s)$ and the one-loop SM functions~\eqref{eq:Pih}. This~applies to the extended Higgs-portal scenarios mentioned in~Section~\ref{sec:conclusions}, such as multi-singlet portals or hidden sectors charged under an unbroken dark gauge group, including strongly coupled ones for which a diagrammatic two-loop calculation is not available. The sum rule~\eqref{eq:sumrule} then relates the coefficient of the IR~logarithms to the reduction of the Higgs-boson couplings, $1 - \kappa_V^2 = \int \! d\hat s \, \rho(\hat s)$, independently of the details of the hidden sector.

\end{appendix}


%

\end{document}